\documentclass[aps,12pt,preprintnumbers,superscriptaddress]{revtex4}
\usepackage{amsmath,amssymb}
\usepackage{color}
\usepackage[toc,page]{appendix}

\renewcommand\d{\partial}

\newcommand\+
{\dagger}
\newcommand\g{\mathrm{g}}

\begin{document}

\preprint{EFI 14-21}
\title{Spacetime Symmetries of the Quantum Hall Effect}
\author{Michael Geracie}
\affiliation{Kadanoff Center for Theoretical Physics, University of 
Chicago, Chicago,
Illinois 60637, USA}
\author{Dam Thanh Son}
\affiliation{Kadanoff Center for Theoretical Physics, University of 
Chicago, Chicago,
Illinois 60637, USA}
\author{Chaolun Wu}
\affiliation{Kadanoff Center for Theoretical Physics, University of 
Chicago, Chicago,
Illinois 60637, USA}
\author{Shao-Feng Wu}
\affiliation{Department of Physics, Shanghai University, Shanghai 
200444, China}
\affiliation{Kadanoff Center for Theoretical Physics, University of 
Chicago, Chicago,
Illinois 60637, USA}

\begin{abstract}
We study the symmetries of non-relativistic systems with an emphasis 
on
applications to the fractional quantum Hall effect. A source for the
energy current of a Galilean system is introduced and the
non-relativistic diffeomorphism invariance studied in previous work is
enhanced to a full spacetime symmetry, allowing us to derive a number
of Ward identities.  These symmetries are smooth in the massless limit
of the lowest Landau level.  We develop a formalism for Newton-Cartan
geometry with torsion to write these Ward identities in a 
covariant form.  Previous results on the connection between Hall
viscosity and Hall conductivity are reproduced.
\end{abstract}

\maketitle

\section{Introduction}

The fractional quantum Hall (FQH)~\cite{Tsui:1982yy,Laughlin:1983}
effect is one of the most difficult problems in condensed matter
physics. In the integer quantum Hall effect, interactions do not play
a large role and one can make much progress by studying the dynamics
of free electrons moving in a uniform magnetic field in the presence
of impurities.  The FQH effect on the other hand, relies crucially on
the interactions of particles within a single Landau level and cannot
be analyzed using perturbative techniques.  The lowest Landau level 
(LLL)
constraint is especially difficult to deal with; the majority of
proposed theoretical schemes break this constraint at some stage.  For
example, in the popular Chern-Simons field theories (in both the
bosonic~\cite{Zhang:1988wy} of fermionic~\cite{Halperin:1992mh}
varieties) the operation of flux attachment mixes states in different
Landau levels.  The consequence of this breaking is that many
physical quantities which should depend only on the Coulomb (or
interaction) energy scale, appear to be sensitive to the cyclotron
energy.  Schemes have been developed within the Chern-Simons field
theory, to evade this unphysical sensitivity to the cyclotron energy,
at the cost of introducing phenomenological elements into the theory.
Some other theoretical approaches have been developed to deal with the
LLL constraint explicitly (see, for example,
Refs.~\cite{Shankar:1997zz,Read:1998dn}), but most have only
limited scope.

This paper proposes a new approach that emphasizes the symmetries of
the LLL.  In recent work we have demonstrated
that non-relativistic particles moving in an external electromagnetic
field possess a far larger degree of symmetry than was previously
realized, namely invariance under arbitrary time-dependent
diffeomorphisms of space~\cite{Son:2013rqa,Son:2005} which may 
further be enlarged to full spacetime diffeomorphism invariance by 
introducing a background source coupled to the energy current 
\cite{Son:2008ye,Karch:2012}.

In section \ref{sec:symmetries} we recap this story, demonstrating how 
diffeomorphism invariance may be obtained by introducing a number of 
different sources.  The source for the
energy density was first introduced by
Luttinger~\cite{Luttinger:1964zz}; including a source coupled to the
energy flux allows for local time reparameterizations.  For specific
interactions, most importantly for a delta-function contact
interaction between bosons,
this action is also Weyl invariant. We then demonstrate
in section \ref{sec:massless} how a regular massless limit may be
taken after a special choice of parity breaking parameters. The
resulting theory contains only particles confined to the LLL. Physical
results of this limit will be considered in upcoming work.

In section \ref{sec:nc ward} we consider the complete set of one-point
Ward identities that follow from non-relativistic diffeomorphism
invariance.  Spatial diffeomorphisms give rise to local momentum
conservation in the presence of external electromagnetic and dilaton
fields whereas temporal diffeomorphisms lead to the work-energy
equation. In trivial backgrounds, these Ward identities were 
considered in Refs.~\cite{Andreev:2013qsa,Karch:2012}. Here we 
present them in their full generality for nonzero spin and $\g$-factor. 
In section~\ref{sec:visc-cond} we rederive the
viscosity-conductivity relations that were first found in
Ref.~\cite{Bradlyn:2012ea}.

Spacetime diffeomorphism invariance can be naturally treated using
the formalism of Newton-Cartan geometry with torsion, which we present
in section~\ref{sec:nc geometry}.  In section \ref{sec:covariant ward}
we present a fully covariant treatment of the one-point Ward
identities. The stress tensor and energy current as traditionally
defined do not transform covariantly under general diffeomorphisms
and need to be modified. We define these covariant currents and derive
their Ward identities. The spacetime transformation properties of the
new covariant stress, charge and energy densities both facilitate
streamlined calculations and place strong constraints on the allowed
response.  Section~\ref{sec:concl} contains concluding remarks while
various technical details are contained in the appendices.

\section{Symmetries}

\label{sec:symmetries} At it's most basic level, the FQH problem is 
that of
particles moving in $2+1$ dimensions in the presence of a magnetic 
field
\begin{equation}  \label{basic action}
S = \int\!d^3x \left( i \psi^\+ D_0 \psi - \frac1{2m} |D_i\psi|^2
-\lambda | \psi |^4 \right) .
\end{equation}
Here $D_\mu = \partial_\mu - i A_\mu$ is the gauge covariant 
derivative and
the theory is gauge invariant
\begin{align}
\psi \rightarrow e^{i \alpha} \psi \qquad A_\mu \rightarrow A_\mu +
\partial_\mu \alpha .
\end{align}
We have chosen a contact interaction for simplicity though more 
general
interactions will be consistent with the symmetries we are about to 
discuss.
Strictly speaking, the contact interaction requires a cutoff to be
well-defined in $2+1$ dimensions due to the logarithmic running of the
coupling constant $\lambda$. In the LLL limit $m\to0$ that we will be
especially interested in, the running of $\lambda$ disappears. We thus 
will
ignore the dependence on the cutoff altogether.

When the magnetic field $B = \epsilon^{ij} \partial_i A_j$ is large 
(here $%
\epsilon^{ij}$ is the antisymmetric symbol with $\epsilon^{12} = 1$), 
the
spectrum is stratified into Landau levels of energy $\frac{B}{m} ( n + 
\frac{%
1}{2} )$ that are well-separated compared to the intra-Landau level 
spacing
(we choose units where $\hbar = c = 1$). Since we are only concerned
with the LLL, we would like to integrate out all states for which $n 
\geq 1$%
. One possible way of doing this is to take the $m \rightarrow 0$ 
limit in
which the higher Landau levels tend to infinity and decouple from the
theory. Unfortunately, this limit is not regular due to the infinite 
shift
in the zero-point energy, but we shall see there is an easy way around 
this.

\subsection{The $\mathrm{g}$-factor}

We will now systematically introduce a number of generalizations to 
the
basic action (\ref{basic action}) that will not affect the physics at 
the
end of the day but are essential for our later analysis. In the 
process we
introduce a number of external probes used to define response 
currents.
Begin with an intrinsic angular momentum parameterized by a g-factor 
$%
\mathrm{g}$
\begin{equation}  \label{basic action 2}
S = \int\!d^3x \left( i \psi^\+ D_0 \psi - \frac1{2m} |D_i\psi|^2 + 
\frac{%
\mathrm{g} B}{4m} | \psi |^2 -\lambda | \psi |^4 \right) .
\end{equation}
In GaAs, this factor is close to zero (there $\mathrm{g}$ is the
product of the Lande g-factor $g^*$~\cite{Weisbuch:1977}) and the
ratio of the band mass $m$ to the bare electron mass: $\mathrm{g}=g^* 
m/m_e
\approx-0.03$), but it is easy to see, at least for constant
$B$, that it's actual value is irrelevant. In this case the new term
merely gives rise to a constant shift to the Hamiltonian, which has no
physical significance.

When $B$ is not uniform, the situation is somewhat more involved but 
not
insurmountable. Notice that $\mathrm{g}$ enters the action in the
combination $A_0 + \frac{\mathrm{g}}{4m} B$. Defining a new electric
potential
\begin{align}  \label{g redefinition}
A^{\prime }_0 = A_0 + \frac{\mathrm{g} - \mathrm{g}^{\prime }}{4m} B
\end{align}
maps the action to itself, but with a new g-factor
\begin{align}
S_\mathrm{g}[A_0 ] = S_{\mathrm{g}^{\prime }} [A^{\prime }_0] .
\end{align}
We may just as well perform calculations with any $\mathrm{g}$ we
like, so long as we use the shifted gauge field $A^{\prime }_0$. In
section \ref{sec:massless} we shall see that when we select
$\mathrm{g}=2$, the LLL is shifted to zero energy even in a non-
uniform
field and curved space. The massless limit is then regular and the
projection onto the LLL proceeds without difficulty. This feature was
exploited in Ref.~\cite{Son:2013rqa} in the construction of an
effective field theory for FQH states.

\subsection{Curvature}

Next introduce a nontrivial background metric $g_{ij}$
\begin{equation}  \label{action with metric}
S = \int\!d^3x \sqrt{g} \left( \frac{i}{2} \psi^\+ 
\overset{\leftrightarrow}{%
D}_0 \psi - \frac{g^{ij}}{2m} D_i\psi^\dagger D_j \psi + 
\frac{\mathrm{g} B}{%
4m} | \psi |^2 -\lambda | \psi |^4 \right)
\end{equation}
where $\psi^\+ \overset{\leftrightarrow}{D}_0 \psi = \psi^\dagger D_0 
\psi -
D_0 \psi^\dagger \psi$. The magnetic field is now $B = 
\varepsilon^{ij}
\partial_i A_j$ where $\varepsilon^{ij} = \frac{1}{\sqrt{g}} 
\epsilon^{ij}$
is the natural spatial volume element associated to the metric. There 
is
some ambiguity in how we choose to couple the theory to geometry; we 
could
for example have included higher curvature terms. These terms would 
change
the equations of motion on curved backgrounds but leave the flat space 
dynamics
unaltered. If at the end of the day one is only interested in flat
space, we may choose the coupling however we like without fear of 
altering
the physics. In the above we have chosen to couple the theory in the 
minimal
way.

If the field $\psi$ has spin $s$, even minimal substitution requires 
the
introduction of a zweibein $e^a_i$ that diagonalizes the metric
\begin{align}
g_{ij} = \delta_{ab} e^a_i e^b_j \qquad e^a_i e^{bi} = \delta^{ab} .
\end{align}
The covariant derivative is then
\begin{align}
D_\mu = \partial_\mu - i A_\mu + i s \omega_\mu
\end{align}
where $\omega_\mu$ is the spin connection
\begin{align}
\omega_0 &= \frac{1}{2} \epsilon_{ab} e^{a j} \partial_0 e^b_j  \notag 
\\
\omega_i &= \frac{1}{2} \epsilon_{ab} e^{aj} \nabla_i e^b_j =\frac12
\epsilon_{ab} e^{aj}\d_i e^b_j - \frac{1}{2} \varepsilon^{jk} \d_j 
g_{ik} .
\end{align}
Here $\nabla_i$ represents the spatial covariant derivative defined by 
$%
g_{ij}$. Under a local rotation of the zweibein by an angle $\theta 
(x)$,
the spin connection transforms as a $U(1)$ gauge field
\begin{align}
\omega_\mu \rightarrow \omega_\mu + \partial_\mu \theta
\end{align}
canceling the spin rotation of the field $\psi \rightarrow e^{- i s 
\theta}
\psi$. We notice that the same minimal coupling to gravity through spin connection was recently used in Ref.~\cite{Cho:2014vfl} to modify the conventional flux attachment procedure to derive the Hall viscosity and Wen-Zee term from the Chern-Simons gauge theories.

Even if one does not care about curved space dynamics and plans to set 
$%
g_{ij} = \delta_{ij}$, introducing a metric is a useful intermediate 
step
for several reasons. First, it gives a natural definition of a 
symmetric
stress tensor as the response of the action to geometric perturbations 
in
the same way that the charge current is a response to electromagnetic
perturbations
\begin{align}  \label{noncovariant currents}
\delta S = \int d^3x \sqrt{g} \left( \frac{1}{2} T^{ij}_\text{nc} 
\delta
g_{ij} + j^\mu_\text{nc} \delta A_\mu \right) ,
\end{align}
as is done in relativity theory. The subscript ``nc" (as in
``non-covariant") is to differentiate this notion of stress from the
spacetime covariant one that we shall introduce later. In the usual 
case $%
\mathrm{g}=s=0$ in flat space, these are the familiar expressions
\begin{align}
j^0_\text{nc} &= | \psi |^2 \qquad \qquad j^i_\text{nc} = - \frac{i}
{2m}
\psi^\dagger \overset{\leftrightarrow}{D^i} \psi  \notag \\
T^{ij}_\text{nc} &= \frac{1}{m}  D^{(i} \psi^\dagger  D^{j)} \psi  + %
\Big( \frac{i}{2} \psi^\dagger \overset{\leftrightarrow}{D}_0 \psi - 
\frac{1%
}{2m} D_k \psi ^\dagger D^k \psi - \lambda |\psi |^4\Big) g^{ij}
\end{align}

The spin is often set to zero in the literature, although for spin 
polarized
electrons in two spatial dimensions, the actual value would be $1/2$.
However, as with $\mathrm{g}$, there is a simple mapping between 
theories of
different spin. Like before, a redefinition
\begin{align}  \label{s redefinition}
A^{\prime }_\mu = A_\mu + (s^{\prime }- s ) \omega_\mu
\end{align}
sends the action to itself, but with $s$ replaced by $s^{\prime }$
\begin{align}
S_{s} [A_\mu ] = S_{s^{\prime }} [A^{\prime }_\mu ] .
\end{align}
In what follows we will find the selection $s=1$ to be particularly
convenient. Note a pure zweibein rotation $A^{\prime }_\mu \rightarrow
A^{\prime }_\mu$, $\omega_\mu \rightarrow \omega_\mu + \partial_\mu 
\theta$
of the new theory now corresponds to a zweibein rotation of angle 
$\theta$
plus a gauge transformation $A_\mu \rightarrow A_\mu - (s^{\prime }- 
s)
\partial_\mu \theta$ of the original theory. Thus in this new picture 
the
field $\psi$ has spin $s^{\prime }$ and local rotation invariance is 
still
manifest.

The redefinitions (\ref{g redefinition}) and (\ref{s redefinition}) 
will
affect the stress and charge current. In appendix 
\ref{sec:translation} we
derive the relationship between the primed and unprimed currents, 
which we
present here in flat space for simplicity. If we imagine doing an 
experiment
on a system with say $\mathrm{g}^{\prime }=s^{\prime }=0$ and change 
to our
preferred values $\mathrm{g}=2$, $s=1$, the new currents are
\begin{align}  \label{translation 1}
j^0_\text{nc} &= j^{\prime 0}_\text{nc} \qquad \qquad j^i_\text{nc} =
j^{\prime i}_\text{nc} + \frac{1}{2m} \epsilon^{ij} \partial_j 
j^{\prime 0}_%
\text{nc}  \notag \\
T^{ij}_\text{nc} &= T^{\prime ij}_\text{nc} - \epsilon^{k ( i} 
\partial_k
j^{\prime j)}_\text{nc} - \frac{1}{2m} \left( B j^{\prime 0}_\text{nc}
\delta^{ij} + \left( \partial^i \partial^j - g^{ij} \partial^2 \right)
j^{\prime 0}_\text{nc} \right) .
\end{align}
The primed currents are to be evaluated at the physical fields 
$E^{\prime
}_i $, $B^{\prime }$, whereas the unprimed currents are at
\begin{align}  \label{translation 2}
E_i &= E^{\prime }_i - \frac{1}{2m} \partial_i B^{\prime }\qquad 
\qquad B =
B^{\prime }.
\end{align}

One of the main reasons for introducing $g_{ij}$ is that it makes the
symmetry of the action more apparent. The index structure makes clear 
that
the theory is invariant under time-independent spatial diffeomorphisms 
$%
\xi^k = \xi^k (\mathbf{x })$
\begin{align}
\delta \psi &= - \xi^k \partial_k \psi  \notag \\
\delta A_\mu &= - \xi^k \partial_k A_\mu - A_k \partial_\mu \xi^k  
\notag \\
\delta g_{ij} &= - \xi^k \partial_k g_{ij} - g_{jk} \partial_i \xi^k -
g_{ik} \partial_j \xi^k .
\end{align}
In Refs.~\cite{Son:2013rqa,Son:2008ye,Son:2005} it was found that this 
invariance may be extended to time-dependent
diffeomorphisms $\xi^k ( t , \mathbf{x })$ by adding a non-covariant 
part to
the transformation of the vector potential
\begin{align}  \label{nonrel spatial diff}
\delta \psi &= - \xi^k \partial_k \psi  \notag \\
\delta A_0 &= - \xi^k \partial_k A_0 - A_k \dot \xi^k + 
\frac{\mathrm{g}-2s}{%
4} \varepsilon^{ij} \partial_i ( g_{jk} \dot \xi^k )  \notag \\
\delta A_i &= - \xi^k \partial_k A_i - A_k \partial_i \xi^k - m 
g_{ij}\dot
\xi^j  \notag \\
\delta g_{ij} &= - \xi^k \partial_k g_{ij} - g_{jk} \partial_i \xi^k -
g_{ik} \partial_j \xi^k
\end{align}
which they called \textit{non-relativistic general coordinate 
invariance}.
The $s$ part has not been considered in previous work. Note that for 
$%
\mathrm{g}=2$, $s=1$, these transformations take a particularly simple 
form.
Taking $s=1$ is mostly a choice made to make the formulas easier to 
work
with, whereas using $\mathrm{g}=2$ is crucial to ensure that the 
regularity
of the $m\to0$ limit.

\subsection{A Source for the Energy Current}

This symmetry may be enlarged further to show that the microscopic
action is not only invariant under time-dependent spatial coordinate
reparameterizations, but completely general changes of coordinates on
spacetime. This allows for a fully spacetime covariant treatment of
non-relativistic physics just as in relativity theory. To do so 
however, we
begin with a seemingly unrelated question: how to define an energy 
current
for our theory.

In general relativity, charged fields couple to both a vector 
potential and
a Lorentzian metric and we can consider the system's response to
infinitesimal variations in these quantities. As in (\ref{noncovariant
currents}) this defines a charge current $j^\mu$ and a stress-energy 
tensor $%
T^{\mu \nu}$ that collects both the energy current $T^{0i}$ and the 
stress $%
T^{ij}$ into a single object. It's well-known that in non-relativistic
physics, we have an independent energy current which we will denote as 
$%
\varepsilon^\mu_\text{nc}$ that is not tied to the stress 
$T^{ij}_\text{nc}$
in any way \cite{Landau:1987}.

A source for the energy current was considered in
Refs.~\cite{Son:2008ye,Karch:2012}. This involves dilaton $\Phi$ and a spatial 
vector $C_i$ in the following way
\begin{equation}  \label{full action}
S = \int\!d^3x \sqrt{g} e^{-\Phi} \left( \frac{i}{2} e^{\Phi} \psi^\+
\overset{\leftrightarrow}{D}_0 \psi - \frac{1}{2m} \Bigl( g^{ij} + 
\frac{i
\mathrm{g}}{2} \varepsilon^{ij} \Bigr) \tilde D_i \psi^\dagger \tilde 
D_j
\psi -\lambda | \psi |^4 \right)
\end{equation}
where $\tilde D_i = D_i + C_i D_0$.  The Hamiltonian now appears in 
the action
with a factor $e^{-\Phi}$ and $\Phi$ is essentially the source 
introduced
by Luttinger~\cite{Luttinger:1964zz}.
Note that we have also collected the
magnetic momentum term into the kinetic term. Upon integration by 
parts,
this merely becomes the $\frac{\mathrm{g} B}{4m} | \psi |^2$ coupling
considered before, plus boundary terms that go as the derivatives of 
the new
fields $\Phi$ and $C_i$.

Finally, for the symmetries we are about to consider to hold, we must 
also
modify the spatial Christoffel symbol to be
\begin{align}
{\Gamma^k}_{ij} = \frac{1}{2} g^{kl} \left( \tilde \partial_i g_{jk} +
\tilde \partial_j g_{ik} - \tilde \partial_k g_{ij} \right) .
\end{align}
In our action this only affects the spin connection
\begin{align}
\omega_i &= \frac{1}{2} \epsilon_{ab} e^{aj} \nabla_i e^b_j =\frac12
\epsilon_{ab} e^{aj}\d_i e^b_j - \frac{1}{2} \varepsilon^{jk} \tilde 
\d_j
g_{ik} .
\end{align}
For $\Phi = C_i = 0$, this is just the action (\ref{action with 
metric}) and
so we have not altered the dynamics in these backgrounds, but we can 
now
define $\varepsilon^\mu_\text{nc}$ via
\begin{align}
\delta S = \int d^3 x \sqrt{g} e^{-\Phi} \left( \frac{1}{2} 
T^{ij}_\text{nc}
\delta g_{ij} + j^\mu_\text{nc} \delta A_\mu + \varepsilon^0_\text{nc}
\delta \Phi + \varepsilon^i_\text{nc} \delta C_i \right).
\end{align}
One might also wish to introduce a source for the momentum current, 
but in a
Galilean invariant theory, the momentum is entirely determined by the 
charge
current, so we do not include any further sources. We will see this in 
section %
\ref{sec:nc ward} and again in section \ref{sec:comparison} where we 
find a
unique way to demonstrate this using Newton-Cartan geometry.

To motivate our placement of $\Phi$ and $C_i$, consider the energy 
current
so defined for $\g=s=0$ and a trivial background $\Phi = C_i = 0$, 
$g_{ij} =
\delta_{ij}$
\begin{align}
\varepsilon^0_\text{nc} &= \frac{1}{2m} D_i \psi^\dagger D^i \psi + 
\lambda
| \psi|^4  \notag \\
\varepsilon^i_\text{nc} &=- \frac{1}{2m} \left( D_0 \psi^\dagger D^i 
\psi +
D^i \psi ^\dagger D_0 \psi \right) .
\end{align}
We immediately recognize $\varepsilon^0_\text{nc}$ as the total energy 
of
the system. One may also check using the equation of motion that the
work-energy equation holds
\begin{align}
\partial_0 \varepsilon^0_\text{nc} + \partial_i 
\varepsilon^i_\text{nc} =
E_i j^i_\text{nc} ,
\end{align}
so it is clear that $\varepsilon^i_\text{nc}$ is indeed the energy 
flux.

The energy current is also altered upon a change of the parity 
breaking
parameters $\mathrm{g}$ and $s$. As before, translating from 
$\mathrm{g}%
^{\prime }=s^{\prime }=0$ to $\mathrm{g}=2$, $s=1$ in the trivial 
background
gives
\begin{align}  \label{translation 3}
\varepsilon^0_\text{nc} &= \varepsilon^{\prime 0}_\text{nc} - \frac{1}
{2}
\epsilon^{ij} \d_i j^{\prime }_{j\text{nc}} - \frac{1}{2m} B j^{\prime 
0}_%
\text{nc}  \notag \\
\varepsilon^i_\text{nc} &= \varepsilon^{\prime i}_\text{nc} + \frac{1}
{2}%
\epsilon^{ij} \partial_0 j^{\prime }{_{j\text{nc}}} - \frac{1}{2m} 
\left( B
j^{\prime i}_\text{nc} + \epsilon^{ij} E_j j^{\prime 0}_\text{nc} 
\right) .
\end{align}
We again refer the reader to appendix \ref{sec:translation} for 
details as
well as the case for general $\mathrm{g}^{\prime }, s^{\prime }$, 
$\mathrm{g}%
, s$ and a curved metric.

\subsection{Spacetime Coordinate Invariance}

Our placement of these new sources does much more than give a 
convenient
definition of the energy current, it allows us to enlarge the group of
spacetime symmetries by properly selecting the transformations of 
$\Phi$ and
$C_i$. The action is invariant under arbitrary spacetime
diffeomorphisms $\xi^\lambda (t , \mathbf{x})$
\begin{align}  \label{noncovariant diff}
\delta \psi &= - \xi^\lambda \partial_\lambda \psi  \notag \\
\delta \Phi &= - \xi^\lambda \partial_\lambda \Phi + \partial_\lambda
\xi^\lambda - \tilde \partial_i \xi^i  \notag \\
\delta C_i &= - \xi^\lambda \partial_\lambda C_i - C_j \tilde 
\partial_i
\xi^j + \tilde \partial_i \xi^0  \notag \\
\delta e^{ai} &= - \xi^\lambda \partial_\lambda e^{ai} + e^{ak} \tilde
\partial_k \xi^i  \notag \\
\delta g^{ij} &= - \xi^\lambda \partial_\lambda g^{ij} + \tilde 
\partial^i
\xi^j + \tilde \partial^j \xi^i  \notag \\
\delta \varepsilon^{ij} &= - \xi^\lambda \partial_\lambda 
\varepsilon^{ij} +
\varepsilon^{ik} \tilde \partial_k \xi^j - \varepsilon^{jk} \tilde
\partial_k \xi^i  \notag \\
\delta A_0 &= - \xi^\lambda \partial_\lambda A_0 - A_\lambda 
\partial_0
\xi^\lambda +\frac{\g- 2 s}{4} \Big( \varepsilon^{ij} \tilde 
\partial_i (
g_{jk} \dot \xi^k ) + \varepsilon^i_{~j} \dot C_i \dot \xi^j \Big)  
\notag \\
\delta A_i &= - \xi^\lambda \partial_\lambda A_i - A_\lambda 
\partial_i
\xi^\lambda - m e^\Phi g_{ij} \dot \xi^j - \frac{\g- 2 s}{4} C_i\Big( 
\varepsilon^{jk} \tilde \partial_j ( g_{kl} \dot \xi^l ) +
\varepsilon^j_{~k} \dot C_j \dot \xi^k \Big)
\end{align}
where $\lambda$ now includes the temporal index 0.

We stress once more that this represents full spacetime coordinate
reparameterization invariance. From (\ref{nonrel spatial diff}) it was 
clear
that the theory was invariant under arbitrary time-dependent 
coordinate
changes on spatial slices. Now we see that the theory is also 
unaffected by
local time reparameterizations $\xi^0 ( t , \mathbf{x})$. In 
particular, we
may choose a new spatial foliation of spacetime. This is another way 
to see
that the new sources are not essential modifications to the physics. 
Even if
$C_i$ was zero initially, in these new slicings, it may not be (for 
example
if the temporal shift is not constant -- $\partial_i \xi^0 \neq 0$), 
just as
the metric $\delta_{ij}$ may not look flat in some other randomly 
chosen
curvilinear coordinate system. Hence we see allowing for nonzero $C_i$ 
is
necessary to make the full spacetime symmetry of the initial theory
manifest, even if we are considering a trivial background that merely
reduces the problem to the original action (\ref{basic action 2}) in 
the end.

There is a minor complication to this story. One could imagine that 
given
some background $C_i$ that there is no slicing where it vanishes. It 
turns
out that a $C_i = 0$ slicing exists if and only if
\begin{align}  \label{GTC}
\varepsilon^{ij} \tilde \partial_i C_j = 0 .
\end{align}
In section \ref{sec:nc geometry} we give this condition a simple 
geometric
interpretation and argue from physical considerations why it must in 
general
be satisfied. Coordinates where $C_i = 0$ are called \textit{global 
time
coordinates} (GTC). Since we shall always assume GTC exist, in what 
follows
we will take $C_i = 0$ without loss of generality, only restoring it 
when
necessary to compute the energy current.

Gauge invariance and spatial diffeomorphisms are not the only local
symmetries of the action (\ref{full action}). For each $\Omega ( t ,%
\mathbf{x})$, the theory also exhibits Weyl invariance
\begin{align}
\delta\psi & =\Omega\psi,\qquad\delta\Phi=2\Omega,\qquad\delta
g_{ij}=-2\Omega g_{ij},\qquad\delta C_{i}=0  \notag \\
\delta A_{0} & =-\frac{1}{2m}\left(1-\frac{\mathrm{g}^{2}}
{4}\right)\left(\frac{1}{\sqrt{g}}\tilde{\partial}_{i}\left(e^{-
\Phi}\sqrt{g}\tilde{\partial}^i\Omega\right)+e^{-\Phi}\dot 
C_{i}\tilde{\partial}^i \Omega\right)  \notag \\
\delta A_{i} & =\frac{\mathrm{g}-2s}{2}\varepsilon_{ij}\tilde{\partial}^j%
\Omega .
\end{align}
This is of course specific to the point interaction $\lambda | \psi
|^4$ where scale invariance is well known to be violated quantum
mechanically~\cite{Thorn:1978kf}. In the massless limit however,
$\lambda$ does not run and the LLL theory is truly conformally
invariant. Note that for $\mathrm{g}=2$, $s=1$, the vector potential
does not transform.

This concludes the complete set of generalizations of the initial
problem that are relevant for this paper. We are now considering
particles of arbitrary spin and g-factor moving in the presence of an
electromagnetic field $A_\mu$, a curved metric $g_{ij}$, a dilaton
$\Phi$ and a spatial vector $C_i$. $\mathrm{g}$ and $s$ may be chosen
at will so long as we remember to translate back to their physical
values using (\ref{translation 1}), (\ref{translation 2}) and
(\ref{translation 3}). In section \ref{sec:nc geometry} we present a
manifestly coordinate invariant treatment of this symmetry from which
the anomalous transformation laws (\ref{noncovariant diff}) follow
naturally. This is the Newton-Cartan geometry first considered in
Ref.~\cite{Son:2013rqa} in the context of the FQH effect. There we shall
find that properly defining the energy current requires a
generalization of this formalism to include nonzero torsion.

\section{The Massless Limit}

\label{sec:massless}

We now perform the massless limit discussed earlier. In GTC the action 
is
\begin{equation}
S = \int\!d^3x \sqrt{g} e^{-\Phi} \left( \frac{i}{2} e^{\Phi} \psi^\+
\overset{\leftrightarrow}{D}_0 \psi - \frac{1}{2m} \big( g^{ij} + 
\frac{i
\mathrm{g}}{2} \varepsilon^{ij} \big) D_i \psi^\dagger D_j \psi -
\lambda |
\psi |^4 \right) ,  \notag
\end{equation}
and the quantum partition function is given by
\begin{align}
Z = \int \mathcal{D }\psi^\dagger \mathcal{D }\psi e^{i S} .
\end{align}
The matrix $\varepsilon^{ij}$ has eigenvalues $\pm i$ and so the value 
$%
\mathrm{g} =2$ is distinguished for the matrix $g^{ij} + i 
\varepsilon^{ij}$
is degenerate. In terms of the zweibein $e^a_i$ we have
\begin{equation}
\varepsilon^{ij} = \epsilon_{ab} e^{ai} e^{bj} .
\end{equation}
The eigenvectors of $\varepsilon^{ij}$ are the chiral basis vectors
\begin{equation}
e_i = \frac1{\sqrt2} (e_i^1 + ie_i^2), \qquad \bar e_i = 
\frac1{\sqrt2}
(e_i^1 - ie_i^2)
\end{equation}
in terms of which we have the convenient formulas
\begin{align}
g^{ij} &= e^i \bar e^j + \bar e^i e^j \qquad \varepsilon^{ij} = i(e^i 
\bar
e^j - \bar e^i e^j) \qquad g^{ij} + i\varepsilon^{ij} = 2\bar e^i e^j 
.
\end{align}

Hence the $\mathrm{g}=2$ action may be written as
\begin{equation}
S = \int d^3 x \sqrt g e^{-\Phi} \left( \frac i2 e^\Phi \psi^\+ 
\overset{%
\leftrightarrow}{D_0} \psi - \frac1{m} (\bar e^i D_i\psi^\+)(e^j 
D_j\psi) -
\lambda |\psi |^4 \right) .
\end{equation}
In flat space, $e^i D_i \psi = D_{\bar z} \psi$ and we see the 
degeneracy
direction corresponds precisely to particles in the LLL. Using a
Hubbard-Stratonovich transformation, we write this as
\begin{equation}
S = \int d^3 x \sqrt g e^{-\Phi} \left( \frac i2 e^\Phi \psi^\+ 
\overset{%
\leftrightarrow}{D_0}\psi - \chi (\bar e^i D_i\psi^\+) -\bar\chi (e^i
D_i\psi) + m \bar\chi\chi - \lambda | \psi |^4 \right) .
\end{equation}
The $m \rightarrow 0$ limit is manifestly regular and the higher 
Landau
levels are now completely trivial to integrate out as $\chi$ and $\bar 
\chi$
simply become Lagrange multipliers enforcing the constraint
\begin{align}  \label{holomorphic}
e^i D_i \psi = 0
\end{align}
which is the curved space equation for the LLL wave function. The 
many-body
problem of particles confined to the LLL thus can be understood as a 
system
of interacting particles with no kinetic energy
\begin{equation}\label{S-LLL}
S = \int d^3 x \sqrt g \left( \frac i2 \psi^\+ 
\overset{\leftrightarrow}{D_0}%
\psi - e^{-\Phi} \lambda | \psi |^4 \right)
\end{equation}
for which path integration is only carried out subject to the 
holomorphic
constraint (\ref{holomorphic}). This theory inherits all the 
symmetries
discussed above. In particular, one may check that both 
Eqs.~(\ref{holomorphic}) and (\ref{S-LLL}) are
preserved by spacetime diffeomorphisms and Weyl transformations.

We note briefly that for $s=1$ the transformation laws are especially 
simple
in the massless limit. In particular, $A_\mu$ is just a one-form
\begin{equation}
\delta A_\mu =-\xi^\lambda\d_\lambda A_\mu - A_\lambda\d_\mu 
\xi^\lambda
\end{equation}
and is unchanged under Weyl transformations.

\section{Non-covariant Ward identities}

\label{sec:nc ward}

In this section we derive the complete set of Ward identities that 
follow
from the symmetries above. We begin with a slight change of viewpoint
however. In section \ref{sec:symmetries} we used a model microscopic 
action $S$
to motivate the introduction of sources and demonstrate the symmetry 
of the
problem. The full quantum dynamics is however determined by the 
effective
action W obtained from integrating out the field $\psi$ and is a 
functional
only of the external fields
\begin{align}
e^{iW} = Z \qquad W = W [ A_\mu , g_{ij} , \Phi , C_i ].
\end{align}
The currents defined from $W$ are then the 1-point expectation values 
of the
microscopic ones defined above. To simply notation we drop the 
brackets $%
\left \langle \right \rangle$ and simply denote them as
\begin{align}
\delta W = \int d^3 x \sqrt{g} e^{-\Phi} \left( \frac{1}{2} 
T^{ij}_\text{nc}
\delta g_{ij} + j^\mu_\text{nc} \delta A_\mu + \varepsilon^0_\text{nc}
\delta \Phi + \varepsilon^i_\text{nc} \delta C_i \right) .
\end{align}

Let's begin with gauge invariance, which is the simplest of the 
symmetries
considered above. The gauge variation of the electromagnetic potential 
is $%
\delta A_\mu = \partial_\mu \alpha$
\begin{align}
\delta W = \int d^3x \sqrt{g} e^{-\Phi} j^\mu_\text{nc} \partial_\mu 
\alpha
= 0 \qquad \implies \qquad 0 = - \int d^3x \partial_\mu \left( 
\sqrt{g}
e^{-\Phi} j^\mu_\text{nc} \right) .
\end{align}
Since $\alpha$ is an arbitrary function of space and time, we conclude 
$%
\partial_\mu ( \sqrt{g} e^{-\Phi} j^\mu_\text{nc} ) = 0$ or
\begin{align}  \label{cons1}
\frac{1}{\sqrt{g}} \partial_0 \left( \sqrt{g} e^{-\Phi} j^0_\text{nc}
\right) + \nabla_i \left( e^{-\Phi} j^i_\text{nc} \right) = 0
\end{align}
which is simply the continuity equation on a curved background with 
dilaton $%
\Phi$.

The remaining Ward identities follow in like manner. Spatial 
diffeomorphisms
$\xi ^{k}$ imply stress conservation
\begin{align}
\frac{e^{\Phi }}{\sqrt{g}}\partial _{0}\left( 
\sqrt{g}\Big(mj_{i\text{nc}}-%
\frac{\mathrm{g}-2s}{4}\varepsilon _{ij}\nabla ^{j}(e^{-\Phi 
}j_{\text{nc}%
}^{0})\Big)\right) & +e^{\Phi }\nabla _{j}\left( e^{-\Phi }
{T_{i}}_{\text{nc}%
}^{j}\right)  \notag \\
& =j_{\text{nc}}^{0}E_{i}+\varepsilon 
_{ij}j_{\text{nc}}^{j}B+\varepsilon _{%
\text{nc}}^{0}\nabla _{i}\Phi .  \label{cons2}
\end{align}%
Note that nonrelativistic diffeomorphism invariance completely 
determines
the momentum current
\begin{equation*}
p_{\text{nc}}^{i}=mj_{\text{nc}}^{i}-\frac{\mathrm{g}-2s}
{4}\varepsilon
^{ij}\nabla _{j}(e^{-\Phi }j_{\text{nc}}^{0})
\end{equation*}%
as mentioned previously. In particular, the mass leads to a momentum 
along
the direction of charge flow while the parity breaking terms give rise 
to an
intrinsic angular momentum density
\begin{equation*}
l=-\frac{\g-2s}{2}e^{-\Phi }j_{\text{nc}}^{0}
\end{equation*}%
as may be seen by computing the flat space total angular momentum 
$L=\int
d^{2}x\epsilon ^{ij}x_{i}p_{j}$. Note that the dilaton exerts an 
external
force on the system just as the electromagnetic field does.

Temporal diffeomorphisms $\xi^0$ result in the work-energy equation
\begin{align}  \label{cons3}
\frac{1}{\sqrt{g}}\partial_0 \big( \sqrt{g} \varepsilon^0_{\text{nc}} 
\big) %
+ e^\Phi \nabla_i ( e^{-\Phi} \varepsilon^i_{\text{nc}} )&= E_i 
j^i_\text{nc}
- \frac{1}{2} T^{ij}_{\text{nc}} \dot g_{ij} .
\end{align}
$E_i j^i$ is the familiar work done by the electric field whereas the
metric term corresponds to the work done on the walls of a volume 
element as
it expands or contracts due to the internal forces of the system. The 
Ward
identities of a system with a conserved particle number thus reproduce 
the full
set of equations of motion for a non-relativistic fluid~\cite{Landau:1987}.

Finally, Weyl invariance gives rise to a generalization of the 
tracelessness
of the stress-energy tensor
\begin{align}
\varepsilon^0_\text{nc} = \frac{1}{2} T^{ij}_\text{nc} g_{ij} + 
\frac{1}{4m}\left(1-\frac{\mathrm{g}^{2}}
{4}\right)e^{\Phi}\nabla_{i}\left[e^{-\Phi}\nabla^{i}\left(e^{-
\Phi}j_{\mathrm{nc}}^{0}\right)\right] - \frac{\mathrm{g} - 2s}{4} 
e^\Phi \varepsilon^{ij}
\nabla_i \left( e^{-\Phi} j_{j \text{nc}} \right) .
\end{align}
Note that the Ward identities take a particularly simple form for the 
LLL
theory: the momentum vanishes and the energy is simply the trace of 
the
stress tensor
\begin{align}
&e^\Phi \nabla_j \left( e^{-\Phi} {T_i}^j_\text{nc} \right) =  j^0_%
\text{nc} E_i + \varepsilon_{ij} j^j_\text{nc} B  + 
\varepsilon^0_{\text{nc}} \nabla_i \Phi , \qquad & 
\varepsilon^0_\text{nc} =
\frac{1}{2} T^{ij}_\text{nc} g_{ij}.
\end{align}

It's worth pointing out that the quantum conservation laws in curved 
space
contain the full information on Ward identities. By taking functional
derivatives of these equations with respect to the sources one obtains
higher order Ward identities which relate the $n$-point correlation 
functions
to $\delta $-function terms involving lower order correlators.

\section{Viscosity-conductivity relation}
\label{sec:visc-cond}

As an illustration, here we will give two Ward identities for two-
point 
functions
and show how they can be used to extract the independent viscosity
coefficients from the conductivities at all frequencies. Our work
provides an alternative field-theory approach to the previous result
\cite{Bradlyn:2012ea} based on a microscopic Hamiltonian and
generalize it to the nonvanishing $\mathrm{g} $-factor and spin.

\subsection{Ward identities on closed time path}

To discuss the real-time response functions, let us invoke the closed
time-path formalism \cite{Schwinger:1961,Keldysh:1964,Chou:1985} and
double Eq.~(\ref{cons2}) on two time branches

\begin{equation}
\partial _{t}\left[ g_{\pm ij}\left( mG_{\pm }^{j}-
\frac{\mathrm{g}-2s}{4}%
\epsilon ^{jk}\partial _{k}\frac{G_{\pm }^{t}}{\sqrt{g_{\pm }}}\right) 
\right] +2\partial _{k}(g_{\pm ij}G_{\pm }^{jk})-\partial _{i}g_{\pm
jk}G_{\pm }^{jk}-E_{\pm i}G_{\pm }^{t}-\varepsilon _{\pm ij}BG_{\pm 
}^{j}=0,
\label{ward}
\end{equation}%
The non-equilibrium current and tensor (density) are defined by%
\begin{equation*}
G_{\pm }^{\mu }\equiv \frac{\delta W}{\delta A_{\pm \mu }},\;G_{\pm
}^{ij}\equiv \frac{\delta W}{\delta g_{\pm ij}}.
\end{equation*}%
The dilaton has been set to zero since it has no effect on the results 
of
this section.

Taking variation of Eq. (\ref{ward}) with respect to sources $A_{l}$ 
and $%
g_{jk}$ on two branches, we obtain four identities. Their linear
combinations give
\begin{equation}
0=\left( m\delta _{ij}\partial _{t}-\epsilon _{ij}B\right) 
G_{\mathrm{ra}%
}^{j,l}(x)-\frac{\mathrm{g}-2s}{4}\delta _{ij}\epsilon ^{jk}\partial
_{t}\partial _{k}G_{\mathrm{ra}}^{t,l}(x)+2\delta _{ij}\partial 
_{k}G_{%
\mathrm{ra}}^{jk,l}(x)+\delta _{i}^{l}G^{t}\partial _{t}\delta (x),
\label{ward1}
\end{equation}%
\begin{eqnarray}
0 &=&\left( m\delta _{l}^{i}\partial _{t}-\epsilon _{\;l}^{i}B\right) 
G_{%
\mathrm{ra}}^{l,jk}(x)-\frac{\mathrm{g}-2s}{4}\epsilon ^{in}\partial
_{t}\partial _{n}G_{\mathrm{ra}}^{t,jk}(x)+2\partial 
_{n}G_{\mathrm{ra}%
}^{in,jk}(x)  \label{ward2} \\
&&+\frac{\mathrm{g}-2s}{8}\epsilon ^{in}G^{t}\partial _{t}\partial
_{n}\delta (x)\delta ^{jk}+\left[ \left( \delta ^{ik}\delta 
_{m}^{j}+\delta
^{ij}\delta _{m}^{k}\right) \partial _{n}\delta (x)-\frac{1}{2}\left( 
\delta
_{m}^{k}\delta _{n}^{j}+\delta _{m}^{j}\delta _{n}^{k}\right) \partial
^{i}\delta (x)\right] G^{mn},  \notag
\end{eqnarray}%
where the correlators $G_{\mathrm{ra}}(x)$ are defined by the second
variation of the generating functional with respect to the sources in
physical presentation \cite{Chou:1985}. It can be split as the 
retarded
Green's function and the contact term%
\begin{equation}
G_{\mathrm{ra}}^{B,A}(x)\equiv \frac{\delta ^{2}W}{\delta 
J_{\mathrm{a}%
}^{B}\left( x\right) \delta J_{\mathrm{r}}^{A}\left( 0\right) 
}=i\theta
(t)\langle \left[ \varphi ^{B}(x),\varphi ^{A}(0)\right] \rangle 
+\langle
\frac{\delta \varphi _{\mathrm{r}}^{B}(x)}{\delta J_{\mathrm{r}}^{A}
(0)}%
\rangle ,  \label{WG}
\end{equation}%
where $\varphi ^{A}$ denotes the conjugate operator of the source 
$J^{A}$.
One can also define the advanced correlators%
\begin{equation*}
G_{\mathrm{ar}}^{B,A}(x)\equiv \frac{\delta ^{2}W}{\delta 
J_{\mathrm{r}%
}^{B}\left( x\right) \delta J_{\mathrm{a}}^{A}\left( 0\right) 
}=i\theta
(-t)\langle \left[ \varphi ^{A}(0),\varphi ^{B}(x)\right] \rangle 
+\langle
\frac{\delta \varphi _{\mathrm{a}}^{B}(x)}{\delta J_{\mathrm{a}}^{A}
(0)}%
\rangle .
\end{equation*}%
Note that after the variation, we have put everything on the 
unperturbed background,
which is assumed as a translationally invariant state with a uniform
magnetic field and a vanishing electric field on the flat spacetime.

Keeping in mind the symmetry%
\begin{equation}
G_{\mathrm{ra}}^{B,A}(x)=G_{\mathrm{ar}}^{A,B}(-x),  \label{ra}
\end{equation}%
and the variation of the continuity equation%
\begin{equation}
\partial _{\mu }G^{\mu ,\nu }(x)=0,\;\partial _{\mu }G^{\mu ,ij}(x)=0,
\label{conn}
\end{equation}%
one can combine Eq. (\ref{ward1}) and Eq. (\ref{ward2}) as

\begin{eqnarray}
&&\left( m\delta _{l}^{n}\partial _{t}-\epsilon 
_{\;l}^{n}B+\frac{\mathrm{g}%
-2s}{4}\epsilon ^{nm}\partial _{m}\partial _{l}\right) \left( m\delta
_{ji}\partial _{t}-\epsilon _{ji}B-\frac{\mathrm{g}-2s}{4}\epsilon
_{i}^{\;k}\partial _{j}\partial _{k}\right) G_{\mathrm{ra}}^{l,j}(x)
\label{read1} \\
&=&4\delta _{ij}\partial _{k}\partial _{m}G_{\mathrm{ra}}^{nm,jk}
(x)+2\delta
_{i}^{n}\partial _{k}\partial _{l}\delta (x)G^{kl}-\left( m\delta
_{i}^{n}\partial _{t}-\epsilon _{\;i}^{n}B\right) \partial _{t}\delta
(x)G^{t}.  \notag
\end{eqnarray}%
In momentum space, this can be recast as a relation at all frequencies
\begin{eqnarray}
&&4G_{\mathrm{ra}}^{j(i,k)l}(\omega )+2\delta ^{lj}G^{ik}  
\label{read1q} \\
&=&\frac{1}{2}m^{2}b^{jm}\left. \frac{\partial ^{2}}{\partial 
q_{i}\partial
q_{k}}G_{\mathrm{ra}}^{m,n}(q)\right\vert _{\vec{q}\rightarrow 
0}b^{nl}-%
\frac{\mathrm{g}-2s}{4}im\left[ \epsilon ^{j(k}G_{\mathrm{ra}%
}^{i),n}(x)b^{nl}+b^{jn}G_{\mathrm{ra}}^{n,(k}(x)\epsilon 
^{i)l}\right] ,
\notag
\end{eqnarray}%
where%
\begin{equation*}
b^{ij}=\omega \delta ^{ij}-i\omega _{c}\epsilon ^{ij},\;\omega 
_{c}=B/m.
\end{equation*}

\subsection{Linear response tensor}

In the following, we study the structure of the correlators,
which allows us to transform Eq. (\ref{read1q}) into a relation 
between the
independent viscosity coefficients and the conductivity.
Define the non-equilibrium current%
\begin{equation}
\langle J^{\mu }(x)\rangle _{\mathrm{r}}\equiv \frac{1}
{\sqrt{g}}\frac{%
\delta W}{\delta A_{\mathrm{a}\mu }(x)},
\end{equation}%
from which we have%
\begin{equation}
\frac{\delta \langle J^{\mu }(x)\rangle _{\mathrm{r}}}{\delta 
A_{\mathrm{r}%
\nu }(0)}=G_{\mathrm{ra}}^{\mu ,\nu }(x).  \label{GJJ1}
\end{equation}%
The deviation in the current from its equilibrium value can be 
formally
expanded in time derivatives%
\begin{equation}
\delta \langle J^{\mu }(x)\rangle _{\mathrm{r}}=-\int d^{3}x^{\prime 
}\sigma
_{1}^{\mu \nu }(x-x^{\prime })\delta A_{\mathrm{r}\nu }(x^{\prime })-
\int
d^{3}x^{\prime }\sigma _{2}^{\mu \nu }(x-x^{\prime })\partial 
_{t}^{\prime
}\delta A_{\mathrm{r}\nu }(x^{\prime })+\cdots .
\end{equation}%
In linear response theory, one usually is interested in the case%
\begin{equation}
\delta \langle J^{i}(x)\rangle _{\mathrm{r}}=\int d^{3}x^{\prime 
}\sigma
^{ij}(x-x^{\prime })\delta E_{\mathrm{r}j}(x^{\prime }).
\end{equation}%
Its variation gives%
\begin{equation}
\frac{\delta \langle J^{i}(x)\rangle _{\mathrm{r}}}{\delta 
A_{\mathrm{r}j}(0)%
}=-\partial _{t}\sigma ^{ij}(x).  \label{GJJ2}
\end{equation}%
Combining Eqs. (\ref{GJJ1}) and (\ref{GJJ2}), the correlator of 
currents can
be expressed as the conductivity tensor
\begin{equation}
G_{\mathrm{ra}}^{i,j}(x)=-\partial _{t}\sigma ^{ij}(x).  \label{GS}
\end{equation}

Similarly, define the non-equilibrium stress tensor%
\begin{equation*}
\langle T^{ij}(x)\rangle _{\mathrm{r}}=\frac{2}{\sqrt{g}}\frac{\delta 
W}{%
\delta g_{\mathrm{a}ij}(x)}.
\end{equation*}%
Then we have%
\begin{equation*}
\frac{\delta \langle T^{ij}(x)\rangle _{\mathrm{r}}}{\delta 
g_{\mathrm{r}%
kl}(0)}=2G_{\mathrm{ra}}^{ij,kl}(x)-\frac{1}{2}\delta ^{kl}\langle
T^{ij}\rangle \delta (x).
\end{equation*}%
Vary the spatial components of the metric and define the
elastic modulus and viscosity tensors by the expansion%
\begin{equation*}
\delta \langle T^{ij}(x)\rangle _{\mathrm{r}}=-\frac{1}{2}\int
d^{3}x^{\prime }\lambda ^{ijkl}(x-x^{\prime })\delta g_{\mathrm{r}%
kl}(x^{\prime })-\frac{1}{2}\int d^{3}x^{\prime }\eta ^{ijkl}(x-
x^{\prime
})\partial _{t}^{\prime }\delta g_{\mathrm{r}kl}(x^{\prime })+\cdots .
\end{equation*}%
In other words%
\begin{equation*}
\frac{\delta \langle T^{ij}(x)\rangle _{\mathrm{r}}}{\delta 
g_{\mathrm{r}%
kl}(0)}=-\frac{1}{2}\lambda ^{ijkl}(x)-\frac{1}{2}\partial _{t}\eta
^{ijkl}(x).
\end{equation*}%
By comparison, we obtain%
\begin{equation}
G_{\mathrm{ra}}^{ij,kl}(x)=\frac{1}{4}\delta ^{kl}\langle 
T^{ij}\rangle
\delta (x)-\frac{1}{4}\lambda ^{ijkl}(x)-\frac{1}{4}\partial _{t}\eta
^{ijkl}(x).  \label{GLY}
\end{equation}

\subsection{Elastic modulus}

The tensor $\lambda^{ijkl}(x)$ is the stress response up to the 
zeroth-order in
time derivatives. However, it is also enough to treat it at zeroth-
order in space
derivatives, since our final goal is to obtain the viscosity at all
frequencies and zero wave number. In other words, we can use the
approximation of the perfect fluid. For a system without magnetic 
field, the
hydrodynamic expansion is given by Ref.~\cite{Gubser:2008sz}%
\begin{equation}
\delta \langle T^{ij}(x)\rangle _{\mathrm{r}}=-\left( P\delta 
^{ik}\delta
^{jl}+\frac{1}{2}\delta ^{ij}\delta ^{kl}\kappa ^{-1}\right) \delta
g_{kl}(x),  \label{hydro1}
\end{equation}%
where $\kappa ^{-1}\equiv -V(\partial P/\partial V)_{S,N}$ is the 
inverse
compressibility. The case of the system with a magnetic field is 
similar.
Since the stress density in a rotationally invariant system
with volume $V$ includes both the pressure $P$ and magnetization 
$M$~\cite{Bradlyn:2012ea}
\begin{equation*}
\langle T^{ij}\rangle =\delta 
^{ij}P_{\mathrm{int}},\;P_{\mathrm{int}}=P-%
\frac{MB}{V},
\end{equation*}%
the constitutive equation at leading order can be written as%
\begin{equation*}
T^{\mu \nu }=\epsilon u^{\mu }u^{\nu }+P_{\mathrm{int}}\left( u^{\mu 
}u^{\nu
}+g^{\mu \nu }\right) .
\end{equation*}%
This result is consistent with the one obtained in 
Refs.~\cite{Jensen:2011xb,Kaminski:2013gca}
both for relativistic and non-relativistic systems, though $B$ is 
taken
there as first order in derivatives.

Consider the energy of the system as $E(N,V,B)=V\varepsilon (\nu ,B)$, 
where
$\nu $ is the filling factor. One can define the internal inverse
compressibility%
\begin{equation}
\kappa _{\mathrm{int}}^{-1}\equiv -V\left. \frac{\partial 
P_{\mathrm{int}}}{%
\partial V}\right\vert _{\nu ,N}=B^{2}\left. \frac{\partial 
^{2}\varepsilon
(\nu ,B)}{\partial B^{2}}\right\vert _{\nu }.
\end{equation}%
Thus, the elastic modulus can be decomposed as
\begin{equation}
\lambda ^{ijkl}(x)=\left[ P_{\mathrm{int}}\left( \delta ^{ik}\delta
^{jl}+\delta ^{il}\delta ^{jk}\right) +\delta ^{ij}\delta ^{kl}\kappa 
_{%
\mathrm{int}}^{-1}\right] \delta (x).  \label{lambda1}
\end{equation}

\subsection{Irreducible decomposition of response tensors}

Any rank-2 tensor can be decomposed as a symmetric trace, a symmetric
traceless and an antisymmetric part, so we have
\begin{equation}
\sigma ^{ij}(x)=\sigma _{L}(x)\delta ^{ij}+\sigma _{T}^{ij}(x)+\sigma
_{H}(x)\epsilon ^{ij},  \label{GJJ3}
\end{equation}%
in which the Hall and longitudinal conductivities%
\begin{equation*}
\sigma _{H}\equiv \frac{1}{2}\left( \sigma ^{12}-\sigma ^{21}\right)
,\;\sigma _{L}\equiv \frac{1}{2}\left( \sigma ^{11}+\sigma 
^{22}\right)
\end{equation*}%
are frequently used in references. Also consider the tensor $\eta
^{ijkl}(x)$ divided into its symmetric and antisymmetric parts in the 
pairs of indices $ij$ and $kl$%
\begin{equation*}
\eta ^{ijkl}(x)=\eta _{S}^{ijkl}(x)+\eta _{A}^{ijkl}(x).
\end{equation*}%
We restrict our interest to the systems with rotational invariance. 
Then the
symmetric part has only two independent components%
\begin{equation}
\eta _{S}^{ijkl}(x)=\zeta (x)\delta ^{ij}\delta ^{kl}+\eta _{sh}
(x)\left(
\delta ^{ik}\delta ^{jl}+\delta ^{il}\delta ^{jk}-\delta ^{ij}\delta
^{kl}\right) ,  \label{yta}
\end{equation}%
and the antisymmetric part has one%
\begin{equation}
\eta _{A}^{ijkl}(x)=\eta ^{H}(x)\left( \delta ^{jk}\epsilon ^{il}-
\delta
^{il}\epsilon ^{kj}\right) .  \label{ytaH}
\end{equation}

\subsection{Viscosity and conductivity}

Now combing Eqs. (\ref{GS}), (\ref{GLY}) and (\ref{lambda1}), we can 
recast
Eq. (\ref{read1q}) as
\begin{eqnarray}
&&\bar{\eta}^{ijkl}(\omega )-\frac{1}{2i\omega }\left( \delta 
^{ij}\delta
^{kl}+\delta ^{kj}\delta ^{il}\right) \kappa _{\mathrm{int}}^{-1}
\label{read3} \\
&=&\frac{m^{2}}{2}b^{jm}\left. \frac{\partial ^{2}\sigma ^{mn}(q)}
{\partial
q_{i}\partial q_{k}}\right\vert _{\vec{q}\rightarrow 0}b^{nl}-
\frac{\mathrm{g%
}-2s}{4}im\left[ b^{nl}\epsilon ^{j(i}\sigma ^{k)n}(\omega 
)+b^{jn}\sigma
^{n(i}(\omega )\epsilon ^{k)l}\right] .  \notag
\end{eqnarray}%
where the \textquotedblleft $-$\textquotedblright\ denotes%
\begin{equation*}
\bar{\eta}^{ijkl}=\frac{1}{2}\left( \eta ^{ijkl}+\eta ^{kjil}\right) 
\text{.}
\end{equation*}%
Note that all the contact terms exactly cancel, up to the term with 
$\kappa
_{\mathrm{int}}^{-1}$.

Plugging Eqs. (\ref{GJJ3}), (\ref{yta}) and (\ref{ytaH}) into Eq. 
(\ref%
{read3}), we can extract respectively the bulk, shear and Hall 
viscosities
from the conductivities at all frequencies%
\begin{equation}
\zeta -\frac{1}{i\omega }\kappa _{\mathrm{int}}^{-1}=\frac{m^{2}}
{2}\left(
\omega ^{2}-\omega _{c}^{2}\right) \left. \frac{\partial ^{2}}
{\partial
q_{1}^{2}}\left[ \sigma ^{11}(q)-\sigma ^{22}(q)\right] \right\vert 
_{\vec{q}%
\rightarrow 0}+\frac{\mathrm{g}-2s}{2}im\left[ i\omega _{c}\sigma
_{L}(\omega )-\omega \sigma _{H}(\omega )\right] ,
\end{equation}%
\begin{equation}
\eta _{sh}(\omega )=\frac{m^{2}}{2}\left. \frac{\partial ^{2}}
{\partial
q_{1}^{2}}\left[ \omega ^{2}\sigma ^{22}(q)+\omega _{c}^{2}\sigma
^{11}(q)+2i\omega _{c}\omega \sigma _{H}(q)\right] \right\vert 
_{\vec{q}%
\rightarrow 0}-\frac{\mathrm{g}-2s}{2}im\left[ i\omega _{c}\sigma
_{L}(\omega )-\omega \sigma _{H}(\omega )\right] ,
\end{equation}%
\begin{equation}
\eta _{H}\left( \omega \right) =\frac{m^{2}}{2}\left. \frac{\partial 
^{2}}{%
\partial q_{1}^{2}}\left[ \left( \omega ^{2}+\omega _{c}^{2}\right) 
\sigma
_{H}(q)-2i\omega \omega _{c}\sigma _{L}(q)\right] \right\vert 
_{\vec{q}%
\rightarrow 0}-\frac{\mathrm{g}-2s}{4}im\left[ \omega \sigma _{L}
(\omega
)+i\omega _{c}\sigma _{H}(\omega )\right] .
\end{equation}%
Note that the above four equations recover Eqs. (4.11-4.14) in 
Ref.~\cite{Bradlyn:2012ea} when $\mathrm{g}-2s=0$.

In the limit of $m\rightarrow 0$, we have the regular identities%
\begin{eqnarray}
&&\bar{\eta}^{ijkl}(\omega )-\frac{1}{2i\omega }\left( \delta 
^{ij}\delta
^{kl}+\delta ^{kj}\delta ^{il}\right) \kappa _{\mathrm{int}}^{-1} \\
&=&-\frac{1}{2}B^{2}\epsilon ^{jm}\epsilon ^{nl}\left. \frac{\partial
^{2}\sigma ^{mn}(q)}{\partial q_{i}\partial q_{k}}\right\vert 
_{\vec{q}%
\rightarrow 0}-\frac{\mathrm{g}-2s}{4}B\left[ \epsilon ^{nl}\epsilon
^{j(i}\sigma ^{k)n}(\omega )+\epsilon ^{jn}\sigma ^{n(i}(\omega 
)\epsilon
^{k)l}\right] ,  \notag
\end{eqnarray}%
and%
\begin{equation}
\zeta -\frac{1}{i\omega }\kappa _{\mathrm{int}}^{-1}=-\frac{1}
{2}B^{2}\left.
\frac{\partial ^{2}}{\partial q_{1}^{2}}\left[ \sigma ^{11}(q)-\sigma
^{22}(q)\right] \right\vert _{\vec{q}\rightarrow 0}-
\frac{\mathrm{g}-2s}{2}%
B\sigma _{L}(\omega ),
\end{equation}%
\begin{equation}
\eta _{sh}(\omega )=\frac{1}{2}B^{2}\left. \frac{\partial ^{2}\sigma 
^{11}(q)%
}{\partial q_{1}^{2}}\right\vert _{\vec{q}\rightarrow 
0}+\frac{\mathrm{g}-2s%
}{2}B\sigma _{L}(\omega ),
\end{equation}%
\begin{equation}\label{visc-cond}
\eta _{H}\left( \omega \right) =\frac{1}{2}B^{2}\left. \frac{\partial
^{2}\sigma _{H}(q)}{\partial q_{1}^{2}}\right\vert 
_{\vec{q}\rightarrow 0}+%
\frac{\mathrm{g}-2s}{4}B\sigma _{H}(\omega ).
\end{equation}
A number of interesting identities of this type were recently found 
for nonzero $\g$ in Ref.~\cite{Gromov:2014}.

\section{Newton-Cartan geometry with torsion}

\label{sec:nc geometry}

The derivation of the Ward identities in the previous sections is
quite straightforward, but the diffeomorphism invariance of the
resulting equations can be verified only by rather cumbersome direct
calculation. We now develop a formalism in which the diffeomorphism
invariance is explicit at each stage of the calculation. That
formalism is a version of Newton-Cartan geometry, which has been
previously applied to the quantum Hall problem was developed
first in the context of non-relativistic gravity by Cartan
\cite{Cartan:1923,Cartan:1924} and may be viewed as the natural 
structure preserved by a gauging of Galilean symmetry~\cite{Kuenzle:1972,Banerjee:2014}. 

This section differs from our previous work in that we consider 
torsionful backgrounds. Torsionful geometries are generally necessary 
in the presence of a nontrivial dilaton field and have for example 
been considered in Ref.~\cite{Christensen:2013}, where it is shown that 
boundary theory corresponding to a $z=2$ Lifschitz spacetime is set in 
a torsionful Newton-Cartan setting. 
We now describe this torsionful
version of Newton-Cartan geometry.

A Newton-Cartan geometry is a manifold endowed with a one-form 
$n_\mu$, a
degenerate metric tensor with upper indices $g^{\mu\nu}$ for which 
$n_\mu$
is a zero eigenvector and a vector $v^\mu$ whose projection onto 
$n_\mu$ is
1
\begin{align}
g^{\mu \nu} n_\mu = 0 \qquad \qquad n_\mu v^\mu = 1.
\end{align}
From $(g,n,v)$ one can uniquely define a metric tensor with lower 
indices $%
g_{\mu\nu}$ by requiring
\begin{equation}
g^{\mu\lambda} g_{\lambda\nu} = \delta^\mu_\nu -v^\mu n_\nu, \quad
g_{\mu\nu} v^\nu =0.
\end{equation}
We define a connection by
\begin{equation}  \label{Gamma}
{\Gamma^\lambda}_{\mu\nu} = v^\lambda \d_\mu n_\nu + \frac12 
g^{\lambda\rho}
( \d_\mu g_{\nu\rho} + \d_\nu g_{\mu\rho} -\d_\rho g_{\mu\nu} ) .
\end{equation}
It is easy to see that under coordinate reparameterizations $%
{\Gamma^\lambda}_{\mu\nu}$ transforms as required for a connection.

In the simplest version of the Newton-Cartan geometry, $n_\mu$ is 
assumed to
be a closed one-form. In this case the connection~(%
\ref{Gamma}) is torsionless: ${\Gamma^\lambda}_{[\mu\nu]}=0$. We shall 
not
assume that this is the case; instead, we only assume the weaker 
condition
\begin{equation}
n \wedge dn = 0 .
\end{equation}
By the Frobenius theorem, $n_\mu$ then locally defines a unique 
spatial slicing to
which $n_\mu$ is normal, giving us a preferred notion of space. This 
condition was also imposed in Ref.~\cite{Christensen:2013} so that 
connection on these slices is the usual, torsionless Riemannian 
connection. However, we note that this is in fact
generally required by the causality of a non-relativistic theory. One 
may show that if $n \wedge dn \neq 0$ at a point $x$, there is a 
neighborhood of $x$ in which every point may be reached by a future 
directed curve (one in which the tangent $u^\mu$ satisfies $n_\mu 
u^\mu > 0$)~\cite{Frankel_book}. In particular, an observer may with 
sufficient speed intersect his own past.

In the case that $dn \neq 0$ the
connection has nonzero torsion
\begin{equation}
{T^\lambda}_{\mu\nu} \equiv 2{\Gamma^\lambda}_{[\mu\nu]} = 2v^\lambda 
\d_{%
[\mu}n_{\nu]} .
\end{equation}
The torsion has the following property: it vanishes when all indices 
are
lowered or raised,
\begin{equation}  \label{torsion-prop}
T_{\lambda\mu\nu}\equiv g_{\lambda\alpha} {T^\alpha}_{\mu\nu} =0 
,\quad
T^{\lambda\mu\nu}\equiv g^{\mu\alpha} g^{\nu\beta} 
{T^\lambda}_{\alpha\beta}
= 0 .
\end{equation}
The first equation comes from $g_{\lambda\alpha}v^\alpha=0$. To see 
the
second equation, one can work in the coordinate system where $n_i=0$.

The connection ${\Gamma^\lambda}_{\mu\nu}$ has some further 
interesting
features. It is compatible with the metric $g^{\mu \nu}$ and with 
$n_\mu$,
\begin{equation}
\nabla_\lambda g^{\mu\nu} = 0, \qquad \nabla_\nu n_\mu = 0 .  
\label{nablan}
\end{equation}
On the other hand, the covariant derivatives of $g_{\mu\nu}$ and 
$v^\mu$ are
nonzero. They can be expressed in terms of the Lie derivative of the 
metric
along $v^\mu$,
\begin{align}
\nabla_\lambda g_{\mu\nu} &= - \tau_{\lambda(\mu} n_{\nu)}, \qquad
\nabla_\nu v^\mu = \frac12 \tau_{\nu\alpha} g^{\alpha\mu} \\
\tau_{\mu\nu} & \equiv \pounds _v g_{\mu\nu} = v^\lambda\d_\lambda
g_{\mu\nu} + g_{\lambda\nu}\d_\mu v^\lambda + g_{\mu\lambda}\d_\nu 
v^\lambda
\end{align}
Using $v^\mu\tau_{\mu\nu}=0$ one can show that
\begin{align}  \label{ids}
& g_{\alpha[\mu}\nabla_{\nu]}v^\alpha = 0 \qquad \nabla_\alpha 
v^{[\mu}
g^{\nu]\alpha} = 0 \qquad g^{\mu\alpha} g^{\nu\beta} \nabla_\lambda
g_{\alpha\beta} =0  \notag \\
& v^\lambda\nabla_\lambda g_{\mu\nu} = 0 \qquad v^\lambda 
\nabla_\lambda
v^\mu = 0.
\end{align}
In fact, it is possible to show that the connection (\ref{Gamma}) is
uniquely determined from equations (\ref{torsion-prop}), 
(\ref{nablan}), and
the first equation in (\ref{ids}). The connection of course also 
defines a
unique volume element by
\begin{align}
\nabla_\rho \varepsilon_{\mu \nu \lambda} = 0.
\end{align}

\subsection{Conservation Laws with Torsion}

\label{sec:conservation}

The way the connection is defined introduces one subtlety which is 
important
for our further discussion. Namely, in a Newton-Cartan theory current
conservation
\begin{equation}
\d_\mu (e^{-\Phi} \sqrt g\, j^\mu) =0
\end{equation}
does not have the familiar form $\nabla_\mu j^\mu=0$, but instead is
\begin{equation}  \label{cons}
(\nabla_\mu-G_\mu)j^\mu =0 \qquad \text{where} \qquad G_\mu=
{T^\nu}_{\nu
\mu} .
\end{equation}
We will find this combination of $\nabla_\mu$ and $G_\mu$ recurring 
often.
This is because the usual formula for integration by parts is modified 
on a
torsionful manifold. Because $\frac{1}{\sqrt{g} e^{-\Phi}} 
\partial_\mu %
\big( \sqrt{g} e^{-\Phi} \big) = {\Gamma^\nu}_{\nu \mu} - {T^\nu}_{\nu 
\mu}$%
, in addition to the usual minus sign, we must also take $\nabla_\mu
\rightarrow \nabla_\mu - G_\mu$ upon an exchange of the derivative.

Furthermore, (\ref{cons}) is consistent with time independence of 
total
charge on a torsionful manifold. By Stokes' theorem
\begin{align}
\int_{\Sigma_1} n_\mu j^\mu - \int_{\Sigma_2} n_\mu j^\mu &= \int
\varepsilon^{\mu \nu \lambda} \partial_\mu j_{\nu \lambda}  \notag \\
&= \int \big( \varepsilon^{\mu \nu \lambda} \nabla_\mu j_{\nu \lambda} 
- {%
T^\nu}_{\nu \mu} j^\mu \big)  \notag \\
&= \int ( \nabla_\mu -G_\mu ) j^\mu =0,
\end{align}
where $\Sigma_1$ and $\Sigma_2$ are spatial slices, $n_\mu j^\mu$ is 
the
charge density and $j_{\mu \nu} = \frac{1}{2} \varepsilon_{\mu \nu 
\lambda}
j^\lambda$ is the dual of $j_\mu$.

\subsection{Coordinate Expressions}

To gain some intuition for the above objects and to connect this 
discussion
with the non-covariant presentation of the previous sections, we 
introduce a
parameterization of the geometry by going into coordinates. In some
coordinate patch, we have without loss of generality
\begin{align}  \label{parameterization}
n_\mu =
\begin{pmatrix}
e^{-\Phi} , & - e^{-\Phi} C_i%
\end{pmatrix}
\qquad v^\mu =
\begin{pmatrix}
e^\Phi (1 + C_j v^j ) \\
e^\Phi v^i%
\end{pmatrix}
.
\end{align}
As we shall see, this is the same $C_i$ introduced previously to 
couple to
the energy current. Because $n \wedge dn = 0$, we may always choose
coordinates where $C_i=0$. Writing out $n \wedge dn = 0$ in 
coordinates
gives (\ref{GTC}), so such coordinates are indeed the global time
coordinates discussed before. From the Newton-Cartan perspective, this
condition is elegant and physically motivated.

However, it is often necessary to work outside of GTC, at least to 
first
order, in order to calculate the energy current. Given (\ref%
{parameterization}), the following coordinate expressions follow
straightforwardly
\begin{align}
g^{\mu \nu} =
\begin{pmatrix}
C^2 & C^j \\
C^i & g^{ij}%
\end{pmatrix}
\qquad g_{\mu \nu} =
\begin{pmatrix}
v^2 & - v_j - v^2 C_j \\
- v_i - v^2 C_i & g_{ij} + v_i C_j + v_j C_i + v^2 C_i C_j%
\end{pmatrix}
.
\end{align}
The anomalous transformation laws (\ref{noncovariant diff}) for 
$\Phi$, $C_i$
and $g_{ij}$ can now be derived from the above expressions as natural
consequences of the covariant transformations
\begin{align}
\delta n_\mu &= - \xi^\lambda \d_\lambda n_\mu - n_\lambda \d_\mu 
\xi^\lambda
& &\delta v^\mu = - \xi^\lambda \d_\lambda v^\mu + v^\lambda 
\d_\lambda
\xi^\mu  \notag \\
\delta g^{\mu \nu} &= - \xi^\lambda \d_\lambda g^{\mu \nu} + \d^\mu 
\xi^\nu
+ \d^\nu \xi^\mu . &
\end{align}

We also have
\begin{align}
\varepsilon_{\mu \nu \lambda} = \sqrt{g} e^{-\Phi} \epsilon_{\mu \nu
\lambda} \qquad \tau_{ij} = e^\Phi \left( \nabla_i v_j + \nabla_j v_i 
+ \dot
g_{ij} \right) .
\end{align}
Here $\epsilon_{\mu \nu \lambda}$ is the antisymmetric symbol with 
$\epsilon_{012} = 1$. The remaining components of $\tau_{\mu \nu}$ are 
specified by the transverse
condition $\tau_{\mu \nu} v^\nu = 0$. We see that $\tau_{\mu \nu}$ is 
a
spacetime covariant form of the fluid shear.

\subsection{The Velocity $v^\protect\mu$ and the Covariant Vector 
Potential}

In the above, $g^{\mu \nu}$ and $n_\mu$ play essential roles with 
clear
physical interpretations. $n_\mu$ gives an absolute notion of space 
via it's
integral submanifolds. $g^{\mu \nu}$ restricts to a Riemannian metric 
on space and
supplies an invariant notion of distance. The ``velocity" vector 
$v^\mu$ on
the other hand is the odd man out, what is it supposed to represent? 
At this
stage in our discussion, it is merely a convenience, an inessential
structure that we use to help define a ``metric inverse" and a 
connection.
It may be selected at will only subject to the constraint $n_\mu v^\mu 
= 1$.
In the presence of a fluid one useful choice is for $v^i$ to simply be 
the
fluid velocity. In this paper however we prefer not to assume anything 
about
the system beyond diffeomorphism invariance, and so do not have any 
preferred
notion for $v^\mu$. We will rather be more concerned with 
demonstrating the $%
v^\mu$ independence of our treatment.

If we were to leave it at that $v^\mu$ may be more trouble than it's 
worth:
whenever it appears we need to worry if the physics depends on an 
arbitrary
choice. However, in the presence of a vector potential $A_\mu$, having 
$%
v^\mu $ around is crucial. In GTC, the vector potential obeys an 
anomalous
transformation law
\begin{align}  \label{lie1}
\delta A_0 &= - \xi^\lambda \partial_\lambda A_0 - A_\lambda 
\partial_0
\xi^\lambda +\frac{\g- 2 s}{4} \Big( \varepsilon^{ij} \tilde 
\partial_i (
g_{jk} \dot \xi^k ) + \varepsilon^i_{~j} \dot C_i \dot \xi^j \Big)  
\notag \\
\delta A_i &= - \xi^\lambda \partial_\lambda A_i - A_\lambda 
\partial_i
\xi^\lambda - m e^\Phi g_{ij} \dot \xi^j - \frac{\g- 2 s}{4} 
\varepsilon^{ij}
C_i \partial_i ( g_{jk} \dot \xi^k )
\end{align}
to first order in $C_i$. Extending the discussion
Ref.~\cite{Son:2013rqa} to arbitrary $\mathrm{g}$ and $s$, we may use 
the
components of $v^\mu$ to define a modified gauge field
\begin{align}  \label{cova}
\tilde A_0 &= A_0 - \frac{1}{2} m e^{\Phi} v^2 - \frac{\mathrm{g} - 2 
s}{4}
\varepsilon^{ij} \big( \tilde \partial_i v_j + \dot C_i v_j \big)  
\notag \\
\tilde A_i &= A_i + m e^\Phi v_i + \frac{1}{2} m e^{\Phi} v^2 C_i + 
\frac{%
\mathrm{g} - 2 s}{4} C_i \varepsilon^{jk}\partial_j v_k
\end{align}
that transforms covariantly under diffeomorphisms
\begin{align}  \label{lie2}
\delta \tilde A_\mu = - \xi^\lambda \d_\lambda \tilde A_\mu - \tilde
A_\lambda \d_\mu \xi^\lambda .
\end{align}

Thus we may use $v^\mu$ to take any invariant effective action phrased 
in
terms of the components $g^{ij}$, $\Phi$, $C_i$ and the vector 
potential $%
A_\mu$ and present it as a functional of only covariant objects
\begin{align}
W[g^{ij} , \Phi , C_i , A_\mu ] = W [ g^{\mu \nu} , n_\mu , \tilde 
A_\mu ,
v^\mu ].
\end{align}
Since the original action carried no $v^i$ dependence, the covariant 
version
must have the following special property: it is invariant under 
changes to $%
v^\mu$ and $\tilde A_\mu$ that leave the physical vector potential 
$A_\mu$
unchanged. In section \ref{sec:comparison}, we use this to provide 
another
demonstration that the momentum of a nonrelativistic system is 
determined by
the charge flow. 

It is easy to now write down our microscopic action in a manifestly 
covariant form using Newton-Cartan geometry
\begin{align}
	S = \int d^3x \sqrt{g} e^{-\Phi} \Big( \frac{i}{2} v^\mu \psi^\dagger 
\overset{\leftrightarrow} D_\mu \psi - \frac{1}{2m} \big( g^{\mu \nu} 
+ \frac{i\g}{2} \varepsilon^{\mu \nu} \big) D_\mu \psi ^\dagger D_\nu 
\psi - \lambda | \psi |^4 \Big) 
\end{align}
where the covariant derivative involves the modified vector potential 
and the Newton-Cartan spin connection
\begin{align}
	\omega_\mu = \frac{1}{2} \epsilon_{ab} e^{a\nu} \nabla_\mu 
e^b_\nu .
\end{align}
 Plugging in the coordinate expressions of the geometry, we find this 
action reduces to the microscopic action considered previously with 
all sources present. Indeed, for $g=s=0$, it was shown in 
Ref.~\cite{Banerjee:2014} that one may generally promote a Galilean 
invariant theory to a diffeomorphism invariant one via the simple 
prescription
\begin{align}
	D_0 \rightarrow v^\mu D_\mu 
	\qquad
	D_\mu \rightarrow e_a^\mu D_\mu
\end{align}
which all that we've done here.

In the LLL limit, the physical vector potential is already a one-form 
and
need not be modified. In this simple case $v^\mu$ is truly unnecessary 
and
can be discarded
\begin{align}
W_\text{LLL}[g^{ij} , \Phi , C_i , A_\mu ] = W_{\text{LLL}} [ g^{\mu 
\nu} ,
n_\mu , A_\mu ].
\end{align}

\section{Covariant Ward identities}

\label{sec:covariant ward}

In section \ref{sec:nc ward} we derived Ward identities by considering
the variation of $A_\mu$, $\Phi$, $C_i$ and $g^{ij}$ under 
nonrelativistic
diffeomorphisms. In this approach, the physical meaning of the 
currents $j_%
\text{nc}^\mu$, $\varepsilon_\text{nc}^\mu$, and $T_{\text{nc}}^{ij}$ 
is
clear, and the resulting Ward identities take the form of the fluid
dynamical equations of motion \cite{Landau:1987}. However from the
Newton-Cartan point of view, the above approach is somewhat unnatural. 
$\Phi$%
, $C_i$, and $g^{ij}$ are merely the components of covariant objects 
$n_\mu$
and $g^{\mu \nu}$ in some choice of coordinates and $A_\mu$ is not 
even a
one-form. Similarly, the above currents do not form spacetime vectors 
and
tensors in an obvious way.

In what follows, we reformulate the previous work in a fully geometric
fashion. We begin with an effective action written as a functional of 
the
geometry and the modified gauge field
\begin{align}
W [ n_\mu , g^{\mu \nu} , v^\mu , \tilde A_\mu]  \label{covariant 
action}
\end{align}
and then define currents $j^\mu$, $\varepsilon^\mu$ and $T^{\mu \nu}$ 
that
transform as spacetime tensors. Covariant Ward identities are derived. 
In
section \ref{sec:comparison} we impose the $v^i$ independence of the 
action
as well as demonstrate the relationship between the covariantly 
defined
currents and the ``nc" currents considered previously.

\subsection{Variation of the Action}

Defining covariant currents requires some care as not all components 
of the
background fields $n_\mu$, $g^{\mu \nu}$ and $v^\mu$ are independent. 
Since
the geometry is constrained to satisfy $n_\mu v^\mu = 1$ and $g^{\mu 
\nu}
n_\nu = 0$, an arbitrary variation is not allowed. Rather, the most 
general
change may be parameterized in terms of an arbitrary $\delta n_\mu$, a
transverse velocity perturbation $\delta u^\mu n_\mu = 0$ and a 
transverse
metric perturbation $\delta h^{\mu \nu} n_\nu = 0$
\begin{align}
\delta n_\mu & & \delta v^\mu = - v^\mu v^\lambda \delta n_\lambda + 
\delta
u^\mu & & \delta g^{\mu \nu} = - v^\mu \delta n^\nu - \delta n^\mu 
v^\nu -
\delta h^{\mu \nu} .
\end{align}
$\delta n_\mu$, $\delta u^\mu$ and $\delta h^{\mu \nu}$ are then 
completely
independent and the currents defined by
\begin{align}
\delta W = \int d^3 x \sqrt{g} e^{-\Phi} \Big( \frac{1}{2} T^{\mu \nu}
\delta h_{\mu \nu}+ j^\mu \delta \tilde A_\mu - \varepsilon^\mu \delta 
n_\mu
- p_\mu \delta u^\mu \Big)  \label{definition currents}
\end{align}
where $T^{\mu \nu}$ and $p_\mu$ are fixed to be transverse
\begin{align}
T^{\mu \nu} n_\nu = 0 \qquad p_\mu v^\mu = 0 .
\end{align}
The only new current in this collection is $p_\mu$. We shall find that 
it is
related to the momentum density of the system and is completely fixed 
by the
$v^\mu$ independence of the effective action.

Under spacetime diffeomorphisms, the background fields change as
\begin{align}
\delta n_\mu &= - n_\lambda \nabla_\mu \xi^\lambda + {T^\lambda}_{\mu 
\nu}
n_\lambda \xi^\nu & &\delta g^{\mu \nu} = \nabla^\mu \xi^\nu + 
\nabla^\nu
\xi^\mu + \big( {{T^\mu}_\lambda}^\nu + {{T^\nu}_\lambda}^\mu \big) %
\xi^\lambda  \notag \\
\delta v^\mu &= - \xi^\lambda \nabla_\lambda v^\mu + v^\lambda
\nabla_\lambda \xi^\mu - {T^\mu}_{\nu \lambda} v^\nu \xi^\lambda & & 
\delta
\tilde A_\mu = - \xi^\lambda \nabla_\lambda \tilde A_\mu - \tilde 
A_\lambda
\nabla_\mu \xi^\lambda + {T^\lambda}_{\mu \nu} \tilde A_\lambda 
\xi^\nu ,
\end{align}
where we have exchanged the coordinate derivatives appearing in 
(\ref{lie1})
and (\ref{lie2}) for covariant derivatives. These immediately give
expressions for $\delta u^\mu = {P^\mu}_\nu \delta v^\nu$ and $\delta 
h^{\mu
\nu} = - {P^\mu}_\lambda {P^\nu}_\rho \delta g^{\lambda \rho}$. Gauge
transformations are of course unchanged
\begin{align}
\delta \tilde A_\mu = \nabla_\mu \alpha .
\end{align}

Proceeding as before, we find the Ward identities corresponding to 
gauge
invariance and diffeomorphism invariance are
\begin{align}
( \nabla_\mu - G_\mu ) j^\mu &= 0  \notag \\
\nabla_\nu ( p_\mu v^\nu ) + p_\nu \nabla_\mu v^\nu + ( \nabla_\nu - 
G_\nu )
{T_\mu}^\nu &- n_\mu ( \nabla_\nu - G_\nu ) \varepsilon^\nu = \tilde 
F_{\mu
\nu} j^\nu - G_{\mu \nu} \varepsilon^\nu  \label{covariant ward}
\end{align}
where we have defined the following notation: $\tilde F_{\mu \nu} = ( 
d
\tilde A )_{\mu \nu}$ is the Newton-Cartan analogue of the 
electromagnetic
field strength and $G_{\mu \nu} = (d n )_{\mu \nu}$ is similarly a
``torsional field strength."

Equations (\ref{covariant ward}) are unfamiliar enough to deserve a 
few
comments. We first observe that current conservation no longer takes 
the
form $\nabla_\mu j^\mu = 0$, but rather $(\nabla_\mu - G_\mu) j^\mu = 
0$ as
discussed in section \ref{sec:conservation}. To bring the second 
equation
into a more enlightening form, we first project it onto spatial slices 
by
raising the index
\begin{align}
\nabla_\nu ( p^\mu v^\nu ) + p_\nu \nabla^\mu v^\nu + ( \nabla_\nu - 
G_\nu )
T^{\mu \nu} = \tilde { F^\mu}_{\nu} j^\nu - {G^\mu}_{\nu} 
\varepsilon^\nu .
\label{stress conservation}
\end{align}
This simply expresses momentum conservation in the presence of 
external
forces. $\tilde {F^\mu}_{\nu} j^\nu$ is of course the usual Lorentz 
force,
but along with the $p_\mu$ terms, also makes contributions to the 
momentum
current due to the modifications necessary to make $\tilde A_\mu$ 
covariant.
For now, merely note that the torsion also exerts a ``Lorentz force,'' 
but
one that couples to the energy current rather than the charge current.

Finally, projecting (\ref{covariant ward}) onto $v^\mu$, we obtain the
Newton-Cartan analogue of the work-energy equation
\begin{align}
(\nabla_\mu - G_\mu ) \varepsilon^\mu = - \tilde F_{\mu \nu} v^\mu 
j^\nu +
G_{\mu \nu} v^\mu \varepsilon^\mu - \frac{1}{2} \tau^{\mu \nu} T_{\mu 
\nu} .
\label{work-energy}
\end{align}
The first two terms on the right hand side represent the work done on 
the
system by the external fields in a frame moving with velocity $v^\mu$. 
In
the case that $v^\mu$ represents a fluid velocity, the physics of the 
final
term is relatively clear: it accounts for energy dissipated due to 
viscous
forces.

\subsection{Comparison with the Noncovariant Approach}

\label{sec:comparison}

Unfortunately, the currents defined in (\ref{definition currents}) 
differ
from the standard currents $T^{ij}_{\text{nc}}$, 
$\varepsilon^\mu_{\text{nc}%
} $ and $j^\mu_{\text{nc}}$ found in the non-covariant Ward 
identities. To
see how, express $\delta n_\mu$, $\delta h^{\mu \nu}$, $\delta u^\mu$ 
and $%
\delta \tilde A_\mu$, in terms of $\delta \Phi$, $\delta C_i$, $\delta
g^{ij} $, $\delta v^i$ and $\delta A_\mu$ and set (\ref{definition 
currents}%
) equal to
\begin{align}  \label{nc def}
\delta W = \int d^3 x \sqrt{g} e^{-\Phi} \Big( \frac{1}{2} 
T^{ij}_{\text{nc}%
} \delta g_{ij} + j^\mu_{\text{nc}} \delta A_\mu+ 
\varepsilon^0_{\text{nc}}
\delta \Phi + \varepsilon^i_{\text{nc}} \delta C_i \Big).
\end{align}
The absence of $\delta v^i$ terms is equivalent to the $v^\mu$ 
independence
of the original action. This procedure then completely fixes $p_\mu$ 
to be
\begin{align}
p_\mu = m j_\mu - \frac{\mathrm{g}-2s}{4} \varepsilon_{\mu \nu} 
\nabla^\nu
\left( n_\lambda j^\lambda \right).
\end{align}
So long as $p_\mu$ takes this value, the identities (\ref{covariant 
ward})
are guaranteed to be independent of changes to $v^\mu$ and $\tilde 
A_\mu$
that leave $A_\mu$ fixed, despite appearances to the contrary. Note 
that
since $j_\mu = g_{\mu \nu} j^\nu$, the $i$th component of $j_{\mu}$ is 
not $%
j_i = g_{ij} j^j$, but rather $j_i - v_i j^0$.

The remaining relationships are
\begin{align}
&T^{ij}_{\text{nc}} = T^{ij} + m e^{\Phi} \big( v^i j^j + v^j j^i - 
j^0 v^i
v^j \big) + \frac{\mathrm{g}-2s}{4} e^\Phi \big( j^0 \Omega g^{ij} - 2
v^{(i} \varepsilon^{j)k} e^\Phi \partial_k ( e^{-\Phi} j^0 ) \big)  
\notag \\
&\varepsilon^0_{\text{nc}} = e^{-\Phi}\varepsilon^0 + m e^{\Phi} \big( 
j_i
v^i - \frac{1}{2} j^0 v^2 \big)  \notag \\
&\varepsilon^i_{\text{nc}} = e^{- \Phi}\varepsilon^i + T^{ij} v_j + 
e^\Phi
v^i v^j p_j + \frac{1}{2} m e^{\Phi} j^i v^2 + \frac{\mathrm{g}-2s}{4} 
\big( %
\Omega j^i + \varepsilon^{ij} v_j e^\Phi \partial_0 ( e^{-\Phi} j^0 ) 
\big)
\notag \\
&j^\mu_{\text{nc}} = j^\mu
\end{align}
where $\Omega = \varepsilon^{ij} \partial_i v_j$. Importantly, note 
that it
is the non-covariant currents that are $v^i$ independent. The
covariant versions will change with the choice of $v^i$, but the above
combinations will not.

For nonvanishing charge density, one convenient choice is $v^i = 
j^i/j^0$.
We then have
\begin{align}
&T^{ij}_{\text{nc}} = m e^{\Phi} j^0 v^i v^j + T^{ij} & 
&j^\mu_{\text{nc}} =
j^\mu  \notag \\
&\varepsilon^0_{\text{nc}} = \frac12 m e^{\Phi} j^0 v^2 +
e^{-\Phi}\varepsilon^0 & &\varepsilon^i_{\text{nc}} = \frac{1}{2} m 
e^{\Phi}
j^0 v^i v^2 + e^{- \Phi}\varepsilon^i + T^{ij} v_j
\end{align}
where we have taken $\mathrm{g}=2$ $s=1$ for simplicity. We thus see 
that in
this frame the covariant currents have a clear physical 
interpretation: $%
T^{ij}$ and $\varepsilon^\mu$ are the \textit{internal} stress and 
energy
currents of the system, that is, the currents that do not arise due to 
the
motion of material from one place to another.

Let's express the covariant Ward identities in terms of the non-
covariant
currents to check their $v^i$ independence. First decompose the
Newton-Cartan field strength $\tilde F_{\mu \nu}$ into the usual
electromagnetic field strength plus the modifications necessary to 
make a
spacetime tensor.
\begin{multline}
\tilde F_{\mu \nu}  = F_{\mu \nu}\\ 
  +
\begin{pmatrix}
0 & m \Big( \partial_0 ( e^\Phi v_j ) + \frac{1}{2} \nabla_j ( e^\Phi 
v^2 ) %
\Big) + \frac{\mathrm{g} - 2 s}{2} \nabla_j \Omega \Big) \\
- m \Big( \partial_0 ( e^\Phi v_i ) + \frac{1}{2} \nabla_i ( e^\Phi 
v^2 ) %
\Big) - \frac{\mathrm{g} - 2 s}{2} \nabla_i \Omega \Big) & m \big( 
\nabla_i
(e^\Phi v_j) - \nabla_j (e^\Phi v_i) \big)%
\end{pmatrix}%
.
\end{multline}

We also require the formula
\begin{align}
{T_\mu}^\nu =
\begin{pmatrix}
0 & - v_k T^{kj} \\
0 & {T_i}^j%
\end{pmatrix}%
,
\end{align}
which follows from the transverseness of the stress tensor. Then 
expanding
out the $0$th and $i$th components of (\ref{covariant ward}), we 
obtain
\begin{align}
\frac{1}{\sqrt{g}} \partial_0 ( \sqrt{g} e^{-\Phi} j^0_\text{nc} ) +
\nabla_i ( e^{-\Phi} j^i_\text{nc} ) &= 0  \notag \\
\frac{1}{\sqrt{g}}\partial_0 \big( \sqrt{g} \varepsilon^0_{\text{nc}} 
\big) %
+ e^\Phi \nabla_i ( e^{-\Phi} \varepsilon^i_{\text{nc}} )&= E_i 
j^i_\text{nc}
- \frac{1}{2} T^{ij}_{\text{nc}} \dot g_{ij}  \notag \\
\frac{e^\Phi}{\sqrt{g}} \partial_0 \Big( \sqrt{g} \big( m j_{i} - 
\frac{%
\mathrm{g} - 2s}{4} \varepsilon_{ij} \nabla^j ( e^{-\Phi} j^0 ) \big) 
\Big) %
&+ e^\Phi \nabla_j ( e^{-\Phi} {T_{\text{nc}i}}^j )  \notag \\
&= j^0_\text{nc} E_i + \varepsilon_{ij} j^j_\text{nc} B + 
\varepsilon^0_{%
\text{nc}} \nabla_i \Phi .  \label{noncovariant ward}
\end{align}
The result is independent of $v^i$ and in perfect agreement with the
non-covariant Ward identities found previously.

\section{Conclusion}
\label{sec:concl}

In this paper we have proposed a new approach to studying the FQH 
effect.
The effort here has been essentially formal and will serve as the 
foundation
of later work where physical consequences are addressed. We've shown 
that by
a special choice of spin and gyromagnetic ratio, a smooth massless 
limit is
obtained and we exactly integrate out all higher Landau levels. This 
choice
can always be made by virtue of a translation formula that tells one 
how to
convert results for one $\mathrm{g}$ and $s$ to any other value.

Furthermore, we have derived the complete set of Ward identities that 
follow from spacetime symmetries in arbitrary backgrounds. These Ward
identities are the usual fluid equations of motion: stress 
conservation and
the work-energy equation, which can be viewed as the consequence of a
spacetime symmetry as in relativity. Finally, a covariant treatment of 
these
Ward identities is then developed that makes that symmetry manifest.

\acknowledgments

We would like to thank A.~G.~Abanov, A.~Gromov K.~Jensen, N.~Read,
P.~Wiegmann for discussions.  This work is supported, in part, by a
Simon Investigator grant from the Simons Foundation, the US DOE grant
No.\ DE-FG02-13ER41958, and the ARO-MURI 63834-PH-MUR grant.
S.-F.~Wu was supported, in part, by NNSFC No. 11275120 and the China
Scholarship Council.

\appendix

\section{From Relativistic to Non-Relativistic Conservation Equations}

In this appendix, we motivate the conservation laws (\ref{cons1}), 
(\ref{cons2}) and (\ref{cons3}) from the relativistic point of view. 
This also makes the physical significance of the dilaton field $\Phi$ 
clearer: it arises as the relativistic lapse function.
We begin with the relativistic continuity equation and conservation of 
stress-energy
\begin{equation}
	\nabla_\mu j^\mu = 0 \qquad
	\nabla_{\mu}{T_{\nu}}^{\mu}=F_{\nu\mu}j^{\mu}
\end{equation}
with the metric ansatz
\begin{equation}
	g_{\mu\nu}=\left(\begin{array}{cc}
		-e^{-2\Phi} & 0\\
		0 & g_{ij}
	\end{array}\right)
\end{equation}
The Christoffel symbol is then
\[
\begin{aligned}
	& {\Gamma^0}_{00}=- \dot \Phi, &  &  &  & 
{\Gamma^0}_{ij}=\frac{1}{2}e^{2\Phi} \dot g_{ij}, &  &  &  & 
{\Gamma^k}_{00}=-e^{-2\Phi}g^{kl}\partial_{l}\Phi\\
 	& {\Gamma^0}_{0i}=-\partial_{i}\Phi, &  &  &  & 
{\Gamma^k}_{0i} = \frac{1}{2}g^{kl} \dot g_{li}, &  &  &  & 
{\Gamma^k}_{ij}=\frac{1}
{2}g^{kl}\left(\partial_{i}g_{lj}+\partial_{j}g_{il}-
\partial_{l}g_{ij}\right) .
\end{aligned}
\]
Plugging this in, we find the continuity equation reads
\begin{align}
	\partial_\mu \left( \sqrt{g} e^{-\Phi} j^\mu \right) = 0
\end{align}
whereas the time and space components of stress-energy conservation 
are
\begin{eqnarray}
	\frac{1}{\sqrt{g}}\partial_0 \left(\sqrt{g}{T_0}^{0}\right)+ 
\nabla_{i} {T_0}^{i}+ e^{-2\Phi} {T_0}^i \partial_i \Phi -\frac{1}
{2}T^{ij} \dot g_{ij} & = & 
F_{0\mu}j^{\mu}\label{eq:Appendix:EnergyConservation}\\
	\frac{1}{\sqrt{g}e^{-\Phi}}\partial_{0}\left( \sqrt{g}e^{-
\Phi}{T_j}^{0}\right)+ e^\Phi \nabla_{i}\left( e^{-\Phi}
{T_j}^{i}\right)+{T_0}^{0}\partial_{j}\Phi & = & 
F_{j\mu}j^{\mu}\label{eq:Appendix:MomentumConservation}
\end{eqnarray}

To bring this into a form
 closer to that which appears in the main text, we define the energy 
density $\varepsilon^0_\text{nc}$, energy
flux $\varepsilon^{i}_\text{nc}$ and momentum density 
$p^{i}_\text{nc}$ as
\[
 {T_0}^{\mu}=- \varepsilon^{\mu}\qquad {T_j}^{0}= p_{j } .
\]
The conservation equations (\ref{eq:Appendix:EnergyConservation})
and (\ref{eq:Appendix:MomentumConservation}) now read
\begin{eqnarray}
	\frac{1}{\sqrt{g}}\partial_{0}\left(\sqrt{g}  \varepsilon^0 
\right)+ e^\Phi \nabla_{i}\left( e^{-\Phi} \varepsilon^{i} \right) & = 
& - F_{0\mu}j^{\mu} - \frac{1}{2}T^{ij} \dot g_{ij}
	\label{eq:Appendix:EnergyConservation_NR}\\
	\frac{e^\Phi}{\sqrt{g}}\partial_{0}\left(\sqrt{g} e^{-\Phi} 
p_{j } \right)+ e^\Phi \nabla_{i}\left( e^{-\Phi} {T_j}^{i}\right) & = 
&  F_{j\mu}j^{\mu} +\varepsilon^0\nabla_{j}\Phi .
	\label{eq:Appendix:MomentumConservation_NR}
\end{eqnarray}
matching our non-covariant ward identities. Of course, the momentum 
and energy currents are not independent in a relativistic theory, but 
they are in the non-relativistic case.

\section{Current Redefinitions -- Noncovariant 
Version}\label{sec:translation}
In section \ref{sec:symmetries} we remarked that how we choose to 
couple the system to curved geometry is largely arbitrary for flat 
space physics. For example, one can imagine adding additional 
curvature terms to the microscopic action. In curved geometry, we 
would of course have different dynamics, but the flat space equations 
of motion would be unchanged. At the same time, non-minimal couplings 
would in general alter the definition of the stress tensor, even in 
flat space.

However, there is another class of modifications that do {\it not} 
affect the dynamics even in a curved background. This freedom has 
great utility: it allows us to choose the parity breaking couplings 
$\g$ and $s$ at will. In particular, we may always choose $\g=2$ and 
$s=1$. The LLL limit then exists and upon taking $m \rightarrow 0$, 
the momentum density vanishes. We now demonstrate how this works in 
detail.

Let's begin with $s$. Consider, as above, a theory of a single field 
$\psi$ with charge $1$ and spin $s$ so that the covariant derivative 
takes the form
\begin{align}
	D_\mu \psi = ( \partial_ \mu - i A_\mu + i s \omega_\mu ) \psi 
.
\end{align}
Assuming that $A_\mu$ and $\omega_\mu$ only appear in the action in 
this way, we may absorb part of $\omega_\mu$ into $A_\mu$
\begin{align}\label{field mod}
	( \partial_ \mu - i A_\mu + i s \omega_\mu ) \psi &= ( 
\partial_ \mu - i  A'_\mu + i  s'  \omega_\mu ) \psi  \nonumber \\
	\text{where ~~}
	A'_\mu &= A_\mu + ( s' - s ) \omega_\mu .
\end{align}
The dynamics of the system is unchanged, but the point of view 
different; we now have a new spin and externally applied 
electromagnetic field.

For simplicity take $\Phi = 0$, $C_i$ = 0. We have two effective 
actions, $S_{s}$ and $S_{s'}$, satisfying
\begin{align}
	W_{s}[ A_\mu , g_{ij}] = W_{s'} [ A_\mu + (s' - s ) \omega_\mu 
, g_{ij}].
\end{align}
Under a metric perturbation, choose a gauge where $\delta e^a_i = 
\frac{1}{2} \delta g_{ij} e^{aj}$ and $\delta e^{ai} =  \frac{1}{2} 
\delta g^{ij} e^a_j$. The perturbed spin connection is then
\begin{align}
	\delta \omega_0 = \frac{1}{4} \varepsilon^j_{~k} \dot g^{ki} 
\delta g_{ij}
	&&\delta \omega_i = - \frac{1}{2} \varepsilon^{jk} \nabla_j 
\delta g_{ik} .
\end{align}
Setting
\begin{align}
	\int d^3x \sqrt{g} e^{-\Phi} \Big( \frac{1}{2} 
T^{ij}_\text{nc} \delta g_{ij} +  j^\mu_\text{nc} \delta A_\mu\Big) = 
\int d^3 x \sqrt{g} e^{-\Phi} \Big( \frac{1}{2}  T'^{ij}_\text{nc} 
\delta g_{ij} +  j'^\mu_\text{nc} \delta A'_\mu\Big) ,
\end{align}
we find a relation between the stress tensors defined in the two 
different pictures
\begin{align}
	j^\mu_\text{nc} = j'^\mu_\text{nc}
	&&
	T^{ij}_\text{nc} = T'^{ij}_\text{nc}  + ( s' - s ) 
\varepsilon^{k ( i} \nabla_k j'^{j)}_\text{nc} + \frac{1}{2} (s' - s) 
\dot g^{k ( i } \varepsilon^{j)}_{~~k} j'^0_\text{nc} .
\end{align}
We are free to choose the spin however we like, so long as we use this 
stress tensor and the modified electromagnetic field (\ref{field 
mod}).

The same procedure allows us to redefine $\g$ as well, though the 
formulas are more cumbersome. Recall the full microscopic action for 
arbitrary $\g$ and $s$ is
\begin{align} \label{micro action}
	S_{gs} &= \int d^3 x \sqrt{g} e^{-\Phi} \Big( \frac{i}{2} 
e^{\Phi} \psi^\dagger \overset{\leftrightarrow} D_0 \psi - \frac{1}
{2m}\big( g^{ij} + \frac{i\g}{2} \varepsilon^{ij} \big) ( \tilde D_i 
\psi )^\dagger ( \tilde D_j \psi ) - \lambda | \psi |^4 \Big) .
\end{align}
We must briefly work outside of GTC, at least to first order, since 
our modifications will affect the energy current. Explicitly 
accounting for all appearances of the vector potential in the 
microscopic action we have
\begin{align}\label{A terms}
	S = \int d^3 x \sqrt{g} e^{-\Phi} \Bigg(  \frac{i}{2} e^\Phi 
\psi^\dagger \overset{\leftrightarrow}{\tilde \partial_0} \psi - 
\frac{1}{2m} \tilde \partial_i \psi^\dagger \tilde \partial^i \psi - 
\frac{i}{2m} \big( \tilde A^i - s \tilde \omega^i+ \frac{\g}{4} 
\varepsilon^{ij} ( \dot C_j - \tilde \partial_j \Phi ) \big) 
\psi^\dagger \overset{\leftrightarrow}{\tilde \partial_i } \psi 
\nonumber \\
	+ e^\Phi \Big( A_0 - s \omega^0+ \frac{\g}{4m} e^{-\Phi} \big( 
F - \varepsilon^{ij} ( \tilde A_i - s \tilde \omega_i ) ( \dot C_j - 
\tilde \partial_j \Phi ) \big) \nonumber \\- \frac{1}{2m} e^{-\Phi} ( 
\tilde A_i - s \tilde \omega_i ) ( \tilde A^i - s \tilde \omega^i ) 
\Big) | \psi |^2 - \lambda | \psi |^4\Bigg)
\end{align}
where for convenience we have defined $F = i \varepsilon^{ij} \tilde 
D_i  \tilde D_j = \Big( B  - \frac{s}{2} R+ \varepsilon^{ij} \big( E_i  
- \frac{s}{2} R_i  \big)  C_j \Big)$. Here $\tilde A_i$ denotes not 
the modified vector potential but $A_i + C_i A_0$. $R$ and $R_i$ are 
the curvature equivalents of the magnetic and electric fields
\begin{align}
	2 \left( \partial_{\mu} \omega_{\nu} - \partial_\nu \omega_\mu 
\right) =
	\begin{pmatrix}
		0 & - R_j \\
		R_i & \varepsilon_{ij} R
	\end{pmatrix} .
\end{align}
$R$ is simply the spatial Ricci scalar and $R_i = \varepsilon^{jk} 
\nabla_j \dot g_{ik}$ measures change in the geometry with time.

We seek a transformation that sends the third and fourth terms of 
(\ref{A terms}) to themselves but with $\g \rightarrow \g'$ and $s 
\rightarrow s'$. The algebra is somewhat prohibitive, but is greatly 
simplified if we only work to leading order in $C_i$ and $\Phi$, which 
gives us enough information to access the currents at least for the 
torsionless case. The transformation
\begin{align}
	A'_0 &= A_0 + (s' - s) \omega_0 + \frac{\g-\g'}{4m} e^{-\Phi} 
F
	- \frac{\g' ( \g' - \g)}{16m} e^{-\Phi} \left( \nabla^i \dot 
C_i - \nabla^2 \Phi \right) \nonumber \\
	A'_i &= A_i + (s' - s) \omega_i + \frac{\g-\g'}{4} 
\varepsilon_i^{~j}( \dot C_j - \tilde \partial_j \Phi ) - \frac{\g-
\g'}{4m} e^{-\Phi} F C_i
\end{align}
does the trick. When $\Phi = 0$, the electric and magnetic fields in 
the new picture are
\begin{align}
	B' &= B + \frac{1}{2} (s'-s) R \nonumber \\
	E'_i &= E_i + \frac{1}{2} ( s' - s ) R_i + \frac{\g - \g'}{4m} 
\nabla_i B .
\end{align}

We have shown then that at the level of the effective action we have
\begin{align}\label{W trans}
	W_{\g s} [ g_{ij} ,\Phi , C_i , A_\mu ] =
	W_{\g's'} [ g_{ij} ,\Phi , C_i , A'_\mu ] .
\end{align}
To relate the one-point correlators in the two conventions, proceed as 
before. Set
\begin{align}
	\int d^3 x \sqrt{g} e^{-\Phi} \Big( & \frac{1}{2} 
T^{ij}_\text{nc} \delta g_{ij} - \varepsilon^\mu_\text{nc} \delta 
n_\mu + j^\mu_\text{nc} \delta A_\mu \Big) \nonumber \\
	&=
	\int d^3 x \sqrt{g} e^{-\Phi} \Big( \frac{1}{2} 
T'^{ij}_\text{nc} \delta g_{ij} - \varepsilon'^\mu_\text{nc} \delta 
n_\mu + j'^\mu_\text{nc} \delta A'_\mu \Big) .
\end{align}

The resulting translation formulas for $\Phi = 0$ are
\begin{align}\label{translate}
	j^0_\text{nc} &= j'^0_\text{nc} \nonumber \\
	j^i_\text{nc} &= j'^i_\text{nc} + \frac{\g-\g'}{4m} 
\varepsilon^{ij} \nabla_j  j'^0_\text{nc}  \nonumber \\
	\varepsilon^0_\text{nc} &= \varepsilon'^0_\text{nc} - 
\frac{\g-\g'}{4} \varepsilon^{ij} \nabla_i  j'_{j\text{nc}}  - 
\frac{\g-\g'}{4m} B j'^0_\text{nc}
	+ \frac{\g' ( \g' - \g )}{16m}  \nabla^2  j'^0_\text{nc}   
\nonumber \\
	\varepsilon^i_\text{nc} &= \varepsilon'^i_\text{nc} - \frac{1}
{2} ( s' - s ) \varepsilon^{ij} \dot g_{jk} j'^k_\text{nc} + \frac{\g-
\g'}{4}\varepsilon^{ij} \partial_0 j'_{j\text{nc}} - \frac{\g-\g'}{4m} 
\left(  B j'^i_\text{nc} + \varepsilon^{ij} \left( E_j - \frac{s}{2} 
R_i \right) j'^0_\text{nc} \right) \nonumber \\
	& - \frac{\g-\g'}{8m} s \left( g^{ij} g^{kl} - g^{il} g^{jl} 
\right) \dot g_{kl} \nabla_j  j'^0_\text{nc}
	- \frac{\g' (\g' - \g )}{16m}   \frac{1}{\sqrt{g}} \partial_0 
\big( \sqrt{g} \nabla^i j'^0_\text{nc} \big)   \nonumber \\
	T^{ij}_\text{nc} &= T'^{ij}_\text{nc}
	+ ( s' - s ) \varepsilon^{k ( i} \nabla_k j'^{j)}_\text{nc} + 
\frac{s' - s}{2} \dot g^{k ( i } {\varepsilon^{j)}}_{k} j'^0_\text{nc} 
\nonumber \\
	&- \frac{\g-\g'}{4m} \Big(   B j'^0_\text{nc} g^{ij} + s \big( 
\nabla^i \nabla^j - g^{ij} \nabla^2 \big)  j'^0_\text{nc}  \Big)
\end{align}

If we merely restricted ourselves to the microscopic action (\ref{full 
action}), the translation formulas are superfluous since we know the 
explicit form of the classical action for all $\g$ and $s$. Rather, 
their power derives from the equality of the full quantum partition 
functions for which we may not have this knowledge. One can imagine 
computing correlation functions for some convenient choice (such as 
$\g=2$, $s=1$ for LLL physics).  $W=W'$ then ensures that regardless 
of that choice we are actually describing the same physics as for the 
true values of $\g$ and $s$ and there is a precise map that can be 
used to determine the physical correlation functions. 
(\ref{translate}) is that map for one-point correlators. One may 
similarly derive a map for two-point correlators, etc. using the 
method above.

\section{Current Redefinitions -- Covariant Version}

The same manipulations above may be carried out for the covariant 
currents as well. To begin, we recall that the microscopic action 
(\ref{full action}) may be written using Newton-Cartan geometry as
\begin{align}
	S = \int d^3x \sqrt{g} e^{-\Phi} \Big( \frac{i}{2} v^\mu \psi^\dagger 
\overset{\leftrightarrow} D_\mu \psi - \frac{1}{2m} \big( g^{\mu \nu} 
+ \frac{i\g}{2} \varepsilon^{\mu \nu} \big) D_\mu \psi ^\dagger D_\nu 
\psi - \lambda | \psi |^4 \Big)
\end{align}
where we have suppressed the volume element and $D_\mu = \nabla_\mu - 
i \tilde A_\mu + i s \tilde \omega_\mu$. Here $\tilde \omega_\mu = 
\frac{1}{2} \epsilon_{ab} e^{a \nu} \nabla_\mu e^b_\nu$ is the spin 
connection associated to a transverse zweibein $g^{\mu \nu} = 
\delta_{ab} e^{a \mu} e^{b \nu}$ and $\tilde A_\mu$ is the modified 
vector potential.

By the same method as above, we find that the substitution
\begin{align}
	\tilde A'_\mu = \tilde A_\mu + ( s' - s ) \tilde \omega_\mu + 
\frac{\g' - \g}{4} \varepsilon_{\mu \nu} G^\nu + \frac{\g-\g'}{4m} 
n_\mu \Big( \tilde F + \frac{\g + \g'}{8} G_\nu G^\nu - \frac{\g'}{4} 
\nabla_\nu G^\nu \Big)
\end{align}
sends the action to itself but with new parity breaking parameters 
$\g'$ and $s'$.
Here $\tilde F = \varepsilon^{\mu \nu} \nabla_\mu ( \tilde A_\nu - s 
\tilde \omega_\nu )$. We may now derive the action with respect to 
$\delta n_\mu$, $\delta h_{\mu \nu}$ and $\delta \tilde A_\mu$ to find 
how our field redefinition has affected the stress, energy and charge 
currents. For brevity, we cite the result only in the flat case $\Phi$ 
= 0, $C_i = 0$, $g_{ij} = \delta_{ij}$.
\begin{align}\label{translation}
	j^\mu &= j'^\mu + \frac{\g-\g'}{4m} \varepsilon^{\mu \nu} 
\nabla_\nu n' \nonumber \\
	\varepsilon^\mu &= \varepsilon'^\mu - \frac{1}{4} ( s' - s ) 
\varepsilon^{\mu \nu} \tau_{\nu \lambda} j'^\lambda - \frac{\g-\g'}{4} 
\tilde B j'^\mu + \frac{\g'-\g}{2} \varepsilon^{\lambda [ \mu} 
\nabla_\nu ( v^{\nu]} j'_\lambda) \nonumber \\
	&- \frac{\g-\g'}{4m} \Big( \varepsilon^{\mu \nu} \tilde F_{\nu 
\lambda} v^\lambda n' - \frac{1}{4} s \bar \tau^{\mu \nu} \nabla_\nu 
n' \Big) - \frac{( \g -\g' ) \g'}{32 m} \Big( \bar \tau^{\mu \nu} 
\nabla_\nu n' - 2 v^\nu (\nabla^\mu \nabla^\nu  -g^{\mu \nu} \nabla^2) 
n' \Big) \nonumber \\
	T^{\mu \nu} &= T'^{\mu \nu} + (s' - s ) \varepsilon^{\lambda ( 
\mu} g^{\nu ) \rho} \nabla_\lambda j'_{\rho}- \frac{\g - \g'}{4m} 
\Big( \tilde B n' g^{\mu \nu} + s ( \nabla^\mu \nabla^\nu - g^{\mu 
\nu} \nabla^2 ) n' \Big)
\end{align}
where $\bar \tau^{\mu \nu}$ is the trace reversed shear and $n =n_\mu 
j^\mu$.



\begin{thebibliography}{99}

\bibitem{Tsui:1982yy} 
  D.~C.~Tsui, H.~L.~Stormer, and A.~C.~Gossard,
  \emph{Two-Dimensional Magnetotransport in the Extreme Quantum 
Limit,}
  Phys.\ Rev.\ Lett.\  {\bf 48}, 1559 (1982).

\bibitem{Laughlin:1983}
  R.~B.~Laughlin,
  \emph{Anomalous Quantum Hall Effect: An Incompressible Quantum Fluid 
  with Fractionally Charged Excitations,}
  Phys.\ Rev.\ Lett.\ 50, 1395 (1983).

\bibitem{Zhang:1988wy} 
  S.~C.~Zhang, T.~H.~Hansson, and S.~Kivelson,
  \emph{Effective-Field-Theory Model for the Fractional Quantum Hall 
Effect,}
  Phys.\ Rev.\ Lett.\  {\bf 62}, 82 (1988).

\bibitem{Halperin:1992mh} 
  B.~I.~Halperin, P.~A.~Lee, and N.~Read,
  \emph{Theory of the half filled Landau level,}
  Phys.\ Rev.\ B {\bf 47}, 7312 (1993).

\bibitem{Shankar:1997zz} 
  R.~Shankar and G.~Murthy,
  \emph{Towards a Field Theory of Fractional Quantum Hall States,}
  Phys.\ Rev.\ Lett.\  {\bf 79}, 4437 (1997).

\bibitem{Read:1998dn} 
  N.~Read,
  \emph{Lowest Landau level theory of the quantum Hall effect: 
  The Fermi liquid-like state,}
  Phys.\ Rev.\ B {\bf 58}, 16262 (1998)
  [cond-mat/9804294].

\bibitem{Son:2013rqa} D.~T.~Son, \emph{Newton-Cartan Geometry and the 
Quantum
Hall Effect,} arXiv:1306.0638 [cond-mat.mes-hall].

\bibitem{Son:2005} 
  D. T. ~Son, and M. ~Wingate,
  \emph{General coordinate invariance and conformal invariance in 
nonrelativistic physics: Unitary Fermi gas,}
  [arXiv:0509786 [cond-mat]].

\bibitem{Son:2008ye} D.~T.~Son,
\emph{Toward an AdS/cold atoms correspondence: A geometric realization 
of the Schr\"odinger symmetry,}
Phys.\ Rev.\ D \textbf{78}, 046003 (2008) [arXiv:0804.3972 [hep-th]].

\bibitem{Karch:2012} 
  S. ~Janiszewski and A. ~Karch,
  \emph{Non-relativistic holography from Ho\v rava gravity,}
  JHEP {\bf 1302}, 123 (2013)
  [arXiv:1211.0005 [hep-th]].

\bibitem{Luttinger:1964zz} 
  J.~M.~Luttinger,
  \emph{Theory of Thermal Transport Coefficients,}
  Phys.\ Rev.\  {\bf 135}, A1505 (1964).

\bibitem{Andreev:2013qsa} 
  O.~Andreev, M.~Haack and S.~Hofmann,
  \emph{On Nonrelativistic Diffeomorphism Invariance,}
  arXiv:1309.7231 [hep-th].

\bibitem{Bradlyn:2012ea} 
  B.~Bradlyn, M.~Goldstein, and N.~Read,
  \emph{Kubo formulas for viscosity: Hall viscosity, Ward identities, 
and the relation with conductivity,}
  Phys.\ Rev.\ B {\bf 86}, 245309 (2012)
  [arXiv:1207.7021 [cond-mat.stat-mech]].

\bibitem{Weisbuch:1977}
  C.~Weisbuch and C.~Hermann,
  \emph{Optical detection of conduction-electron spin resonance in 
GaAs,
  Ga$_{1-x}$In$_x$As, and Ga$_{1-x}$Al$_x$As,}
  Phys.\ Rev.\ B {\bf 15}, 816 (1977).
	
\bibitem{Cho:2014vfl} 
  G.~Y.~Cho, Y.~You and E.~Fradkin,
  \emph{The Geometry of Fractional Quantum Hall Fluids,}
  arXiv:1406.2700 [cond-mat.str-el].

\bibitem{Landau:1987} L.~D.~Landau and E.~M.~Lifshitz \emph{Course of
Theoretical Physics, Vol. 6: Fluid Mechanics,} Pergamon\ Press, Oxford 
U.K.
(1987).

\bibitem{Thorn:1978kf} 
  C.~B.~Thorn,
  \emph{Quark confinement in the infinite momentum frame,}
  Phys.\ Rev.\ D {\bf 19}, 639 (1979).

nonrelativistic
Ann. Phys.\ (N.Y.) \textbf{321}, 197 (2006) [cond-mat/0509786].

\bibitem{Schwinger:1961} J. Schwinger, \emph{Brownian Motion of a 
Quantum
Oscillator,}\ J. Math. Phys. \textbf{2,} 407 (1961).

\bibitem{Keldysh:1964}
L. V. Keldysh, \emph{%
Diagram Technique For Nonequilibrium Processes,}\ Zh. Eksp. Teor. Fiz.
\textbf{47,} 1515 (1964) [Sov. Phys. JETP \textbf{20,} 1018 (1965)].

\bibitem{Chou:1985} See a review: K. C. Chou, Z. B. Su, B. L. Hao, and 
L. Yu,
\emph{Equilibrium and nonequilibrium formalisms made unified,} Phys. 
Report.
\textbf{118,} 1 (1985).

\bibitem{Gubser:2008sz} 
  S.~S.~Gubser, S.~S.~Pufu, and F.~D.~Rocha,
  \emph{Bulk viscosity of strongly coupled plasmas with holographic 
duals,}
  JHEP {\bf 0808}, 085 (2008)
  [arXiv:0806.0407 [hep-th]].

\bibitem{Jensen:2011xb} 
  K.~Jensen, M.~Kaminski, P.~Kovtun, R.~Meyer, A.~Ritz, and A.~Yarom,
  \emph{Parity-Violating Hydrodynamics in 2+1 Dimensions,}
  JHEP {\bf 1205}, 102 (2012)
  [arXiv:1112.4498 [hep-th]].

\bibitem{Kaminski:2013gca} 
  M.~Kaminski and S.~Moroz,
  \emph{Non-Relativistic Parity-Violating Hydrodynamics in Two Spatial 
Dimensions,}
  Phys.\ Rev.\ B {\bf 89}, 115418 (2014)
  [arXiv:1310.8305 [cond-mat.mes-hall]].

\bibitem{Gromov:2014} 
  A. ~Gromov and A. ~Abanov,
  \emph{Density-curvature response and gravitational anomaly,}
  [arXiv:1403.5809 [cond-mat.str-el]].

\bibitem{Cartan:1923} E.~ Cartan, \emph{Sur les vari\'et\'es \`a 
connexion
affine et la th\'eorie de la relativit\'e g\'en\'eralis\'ee 
(premi\`ere
partie),} Ann.\ Sci.\ \'Ecole Norm.\ Sup. \textbf{40}, 325 (1923).

\bibitem{Cartan:1924} E.~ Cartan, \emph{Sur les vari\'et\'es \`a 
connexion
affine et la th\'eorie de la relativit\'e g\'en\'eralis\'ee 
(premi\`ere
partie) (Suite),} Ann.\ Sci.\ \'Ecole Norm.\ Sup. \textbf{41}, 1 
(1924).

\bibitem{Kuenzle:1972} 
  H.~ P.~ Kuenzle,
  {\em Galilei and Lorentz structures on space-time - comparison of 
the corresponding geometry and physics,}
  Ann.\ Inst.\ Henri\ Poincaré {\bf 17} (1972) 337.

\bibitem{Banerjee:2014} 
  R. ~Banerjee, A. ~Mitra and P. ~Mukherjee,
  \emph{A new formulation of non-relativistic diffeomorphism 
invariance,}
  [arXiv:1404.4491 [gr-qc]].

\bibitem{Christensen:2013} 
  M. ~Christensen, J. ~Hartong, N. A. Obers and B. ~Rollier,
	\emph{Torsional Newton-Cartan Geometry and Lifshitz Holography,}
  Phys.\ Rev.\ D {\bf 89}, 061901 (2014)
  [arXiv:1311.4794 [hep-th]]; 
  \emph{Boundary Stress-Energy Tensor and Newton-Cartan Geometry in 
Lifshitz Holography,}
  JHEP {\bf 1401}, 057 (2014)
  [arXiv:1311.6471 [hep-th]].

\bibitem{Frankel_book}
See, e.g., T.~Frankel, 
\emph{The Geometry of Physics: An Introduction,} 3rd ed., Cambridge
University Press, 2011, Section 3.6d. 




\end{thebibliography}
\end{document}